\documentclass{article}

\usepackage{arxiv}

\usepackage[utf8]{inputenc} 
\usepackage[T1]{fontenc}    
\usepackage{hyperref}       
\usepackage{url}            
\usepackage{booktabs}       
\usepackage{amsfonts}       
\usepackage{nicefrac}       
\usepackage{microtype}      
\usepackage{lipsum}		
\usepackage{graphicx}
\usepackage[super,numbers]{natbib}
\usepackage{doi}
\usepackage{amsthm}
\usepackage{amsmath}
\usepackage{threeparttablex}
\usepackage{dcolumn}
\newtheorem{assumption}{Assumption}
\newtheorem{proposition}{Proposition}

\title{Constructing a T-test for Value Function Comparison of Individualized Treatment Regimes in the Presence of Multiple Imputation for Missing Data}

\author{{Minxin Lu} \\
	Department of Biostatistics\\
	University of North Carolina at Chapel Hill\\
	North Carolina, U.S.A. \\
	\texttt{minxin.lu@unc.edu} \\
	\And
	{Annie Green Howard} \\
	Department of Biostatistics, Carolina Population Center\\
	University of North Carolina at Chapel Hill\\
	North Carolina, U.S.A. \\
	\texttt{aghoward@email.unc.edu} \\
	\AND
	Penny Gordon-Larsen \\
	Department of Nutrition, Carolina Population Center\\
	University of North Carolina at Chapel Hill \\
        North Carolina, U.S.A. \\
	\texttt{pglarsen@unc.edu} \\
	\And
	Katie A. Meyer \\
	Department of Nutrition \\
	University of North Carolina at Chapel Hill \\
        North Carolina, U.S.A. \\
	\texttt{ktmeyer@email.unc.edu} \\
        \And
	Hsiao-Chuan Tien \\
	Carolina Population Center \\
	University of North Carolina at Chapel Hill \\
        North Carolina, U.S.A. \\
	\texttt{htien@unc.edu} \\
        \And
	Shufa Du \\
	Department of Nutrition, Carolina Population Center \\
	University of North Carolina at Chapel Hill \\
        North Carolina, U.S.A. \\
	\texttt{dushufa@email.unc.edu} \\
        \And
	Huijun Wang \\
	National Institute for Nutrition and Health \\
	Chinese Center for Disease Control and Prevention \\
        Beijing, China \\
	\texttt{wanghj@ninh.chinacdc.cn} \\
        \And
	Bing Zhang \\
	National Institute for Nutrition and Health \\
	Chinese Center for Disease Control and Prevention \\
        Beijing, China \\
	\texttt{zhangbing@chinacdc.cn} \\
        \And
        Michael R. Kosorok\\
	Department of Biostatistics\\
	University of North Carolina at Chapel Hill\\
	North Carolina, U.S.A. \\
        \texttt{kosorok@bios.unc.edu}
}

\renewcommand{\shorttitle}{Lu et al}

\begin{document}
\maketitle

\begin{abstract}
	Optimal individualized treatment decision-making has improved health outcomes in recent years. The value function is commonly used to evaluate the goodness of an individualized treatment decision rule. Despite recent advances, comparing value functions between different treatment decision rules or constructing confidence intervals around value functions remains difficult. We propose a t-test based method applied to a test set that generates valid p-values to compare value functions between a given pair of treatment decision rules when some of the data are missing. We demonstrate the ease in use of this method and evaluate its performance via simulation studies and apply it to the China Health and Nutrition Survey data.
\end{abstract}


\section{Introduction}
The response to a particular treatment can vary among individuals due to the influence of their unique characteristics. This variability in treatment effect is particularly relevant in healthcare, where individualized treatment regimes (ITRs) are employed. ITRs tailor treatments and preventive approaches for individuals based on factors such as their socioeconomic status, environment, lifestyle choices, and medical conditions. The application of ITRs extends beyond healthcare and medical treatment assignments, encompassing precision nutrition and behavioral interventions as well.

In these applications, decision-makers may seek to compare the effectiveness of an ITR against, for example, the observed treatment assignment, a one-size-fits-all treatment approach, or another ITR. This comparison is valuable for both patients and healthcare providers in directing their efforts toward the most effective treatment option. For example, individuals with or at risk for hypertension are generally advised to increase their physical activity. The American Heart Association, based on the Physical Activity Guidelines for Americans \citep{piercy2018physical}, recommends that adults engage in  2.5-5 hours of moderate-intensity physical activity per week; although it is noted that additional health benefits can be achieved by achieving beyond 5 hours per week \citep{PAguideline}. Nevertheless, this physical activity recommendation can be challenging to meet for many people. In addition, it is possible that this level of physical activity may not be beneficial for all individuals \citep{bouchard2001individual}.

Thus, it is important to examine whether physical activity, especially at high recommended levels, which might be challenging to implement, uniformly benefits hypertension prevention and management across all demographic groups. This may be particularly relevant in countries in which the prevalence of anti-hypertensive medications is low, such as China. We aim to use ITRs to differentiate between subpopulations that experience incremental benefits from increasing their weekly physical activity to more (versus less or equal) than 5 hours per week and directly compare the ITR generated to population-level recommendations. For subpopulations who are unable to adhere to recommended activity levels, there are potential significant benefits conferred by focusing on alternative behavior interventions, such as dietary modification. Moreover, there are various methods for estimating the ITRs, such as Q-learning and D-learning \citep{qian2011performance,tian2014simple}. Direct comparisons of these ITRs are valuable for determining whether one ITR results in a superior population-level improvement, as compared to another ITR, or to a non-individualized treatment.

Observational data is frequently encountered in research studies and serves as a vital data source for deriving ITRs. Observational epidemiologic studies have several potential strengths, including increased representation of variability in the target population, through large sample sizes and sampling strategies to increase population representation. However, observational data generally lack balance in treatment assignments. To address this confounding, the average outcome for the population, assuming adherence to a specific treatment regime, is weighted by inverse individual propensity scores, and this weighted outcome is defined as the value function. Researchers have commonly employed the value function to evaluate the performance of an ITR derived from observational cohort data. The presence of missing data constitutes a common challenge encountered within randomized trial datasets as well as observational datasets. While large observation data potentially provides more heterogeneity than smaller, more controlled studies, observational studies are often more susceptible to missing data problems. One efficacious approach involves the application of multiple imputation \cite{sterne2009multiple} whereby multiple plausible imputed datasets are generated. Each imputed dataset entails the substitution of missing values with imputed values derived from a predictive model. Subsequently, standard statistical methods are applied to each of the imputed datasets separately. This process yields a collection of results, and the observed variability among these results across the imputed datasets reflects the inherent uncertainty associated with missing values. Multiple imputation allows comprehensive assessment of the robustness of the study findings in the presence of missing data and has the potential to enhance reliability and generalizability of the research outcomes.

Various approaches have been employed to estimate the variance of a value function for ITRs, such as jackknife\citep{jiang2021precision}, cross-validation \citep{cui2017tree, zhao2012estimating}, Q-function model-based approaches \citep{shi2022statistical}, and bootstrapping \citep{efron1992bootstrap}. Each is explained below. However, these approaches tend to become intricate when using multiple imputation and are not intended for directly comparing the value functions of a pair of ITRs, such as estimating the variance around the difference between two value functions. Jiang \textit{et al}\cite{jiang2021precision}. employed the jackknife or leave-one-out cross-validation method to estimate the variance of the value function and compared 24 individualized treatment regimes derived from 24 machine learning models. This jackknife approach consistently estimates the variance around a value function estimate and requires only weak assumptions such as requiring the samples to be independent and identically distributed, and that as the sample size becomes larger, the decision rule estimated from n-1 individuals converges to the decision rule estimated from n individuals. However, it is a time-consuming implementation and better suited for small datasets. Cross-validation provides an alternative means to estimate variance. Cui \textit{et al}.\cite{cui2017tree} and Zhao \textit{et al}. \cite{zhao2012estimating} used cross-validation in their paper. The data is divided into $K$ folds, and the ITR was estimated based on the $K-1$ folds and the value function was evaluated based on the remaining fold. The process is repeated $K$ times to get $K$ estimates of value functions and then the variance of value function is calculated. A common choice of $K$ is $K=10$ which is called 10-fold cross-validation. This method is less time-consuming than the jackknife, though it has a tendency to understate the actual variance. This is because each data point is utilized in both the training and testing sets, leading to a correlation among the accuracy measures of each fold. \citep{bates2021cross}  Shi \textit{et al}.\cite{shi2022statistical} have developed a method for constructing the statistical inference of a policy’s value function for reinforcement learning when either the number of decision points or the sample trajectories diverge to infinity. Their approach involves utilizing the sieve method to approximate the Q- function and employs “SequentiAl Value Evaluation” to split data and iteratively find the optimal policy and the value function estimate. Their method constructs a valid confidence interval around the value function estimate that achieves the nominal coverage. However, it requires estimation of the Q-function and relies on the correctness of the Q-function model. Bootstrapping provides an alternative approach to compute the variance of the value function for an ITR. The standard n-out-of-n bootstrap method \cite{efron1992bootstrap} involves randomly selecting n observations with replacements from the original data with n observations to create new datasets, which are then repeatedly generated 500 or 1000 times. Then, methods for finding the value function of the ITR are applied to each sequence, resulting in slightly different value functions. Subsequently, the variance and confidence interval of the ITR's value function are developed. Variations of the standard n-out-of-n bootstrap method include double bootstrap\cite{chakraborty2010inference}, adaptive bootstrap\cite{laber2011adaptive}, and m-out-of-n bootstrap \cite{chakraborty2013inference}. All bootstrapping methods require repeating operations on the data, usually 500 times or more, making the approach computationally intensive.

In this paper, we present a new method that enables the direct comparison of any two ITRs via the value function and also provides a t-test-based p-value for the significance of observed differences. Our approach addresses the shortcomings described above in that: 1) it is less time-consuming to implement than the jackknife or bootstrapping;  2) it circumvents the problem of correlation in each fold experienced by cross-validation; and 3) it does not rely on the estimation of the Q-function. Our method is suitable for both observational studies and clinical trials. Moreover, our method enables deriving variance estimates from multiple imputed datasets without the need for additional replication or bootstrapping, making it particularly advantageous and efficient when dealing with missing data. Specifically, our approach provides a valid estimate of the variability surrounding both the value function itself and the difference between the two value functions. With our approach, it is possible to assess whether the optimal ITR significantly outperforms the one-size-fits-all approach or another ITR estimated using a different method. Furthermore, our method is characterized by its ease of understanding and implementation, while providing theoretical guarantees for estimator consistency and inference validity, including variance and confidence interval calculations. For illustrative purposes, we use two models, Q-learning \citep{qian2011performance} and D-learning \citep{tian2014simple} for ITR comparison, although more sophisticated models can also be incorporated.

The following outlines the structure of this paper. Section 2 presents the introduction of our method. In section 3, we showcase the application of our method through simulation studies. Section 4 delves into the implementation of our method on a real-world dataset. Finally, in Section 5, we discuss the advantages and limitations of our approach.

\section{Methods}\label{sec2}

In this section, we detail the methodology for calculating the t-test based p-value to assess the difference between two individualized treatment rule (ITR) value functions. We initiate our discussion by defining the framework and introducing relevant notations. Subsequently, we present Proposition \ref{prop:vvar}, which outlines the limiting distribution of the variance estimation for the estimated value function. We then proceed to establish the limiting distribution of the difference between two value functions as delineated in Proposition \ref{prop:vvardiff}, and describe the derivation of the corresponding t-test statistics for pairwise value function comparisons in Equation \ref{eq:ttest}. Lastly, we discuss the adaptation of this variance estimation and t-test statistics to accommodate datasets with missing entries via multiple imputation in Equation \ref{eq:ttestmi}.

We assume that the dataset is divided into a training set that estimates the decision rule and a test set that evaluates this rule on new data, with $n$ and $m$ individuals respectively. The training data is independent of the test data. The individual index is denoted by $i$, where $X_i$ represents the covariate vector, $A_i$ the binary treatment assignment, and $Y_i$ the outcome. Let $\hat{d}_{1,n}$ and $\hat{d}_{2,n}$ be two decision rules we want to compare. Each $\hat{d}_n$ is a map from covariates to treatments. The $\hat{d}_{1,n}$ and $\hat{d}_{2,n}$ can be zero-ordered decision rules (e.g. assign everyone the same treatment), or ITRs as a function of the covariates based on models from the training set (e.g. assign everyone who is younger than 40 years old one treatment, and assign placebo otherwise). The variable $n$ indicates that the rules are estimated using training data, and in case of missing data, decision rules are derived from single or multiple imputation of the training data. The main contribution of our new method provide the estimates for the following:

\begin{itemize}
\item The value function of the decision rule $d_{j,n}$ on the test set $V_{m}(d_{j,n})=E(Y_{m}|A=\hat{d}_{j,n}(x))$, for $j =1,2$
    \item  The variance (and 95\% confidence interval) associated with the value function $Var(V_{m}(d_{j,n}))$
    \item The difference between value functions of two decision rules $V_{m}(d_{1,n})-V_{m}(d_{2,n})$
    \item The variance (and 95\% confidence interval) of the difference between the value functions of two decision rules $Var(V_{m}(d_{1,n})-V_{m}(d_{2,n}))$
    \item The p-value of the t-test for the significance of the difference between the value functions of two decision rules $V_{m}(d_{1,n})-V_{m}(d_{2,n})$
\end{itemize}

\subsection{Propensity Score}
Although randomized trial data are ideal for estimating the optimal ITR, many researchers only have access to observational data. 
When using observational data, the confounding effect is a main concern and can be mitigated by propensity score modeling \citep{austin2011introduction}. The propensity score, denoted as $\pi(A|X) = P(A|X)$, models the probability of receiving treatment $A$ given covariate vector $X$. By weighting each individual with the inverse propensity score, the observational data can approximate the data from a randomized trial. Let $\hat{\pi}_n(a|x) = \pi_n(a|x,\hat{\theta}_n)$ be the estimated propensity score from the training set, and $\pi_0(a|x) = \pi_n(a|x,\theta_0)$ be the true propensity score, where $\hat{\theta}_n$ and $\theta_0$ denote the estimated and the true parameters for the propensity score model. 

\begin{assumption} \label{proposition_propen}
Let $\phi_0(A,X)$ be a vector of functions of $A$ and $X$, such that the following equation is satisfied:
\begin{equation}
    \sqrt{n}(\pi(a|x,\hat{\theta}_n)-\pi(a|x,\theta_0))=\sqrt{n}(\hat{\theta}_n-\theta_0)^T\phi_0(a,x)+o_p(1).
\end{equation}
 \end{assumption}

\begin{assumption}

\begin{equation}
    \sqrt{n}(\hat{\theta}_n-\theta_0) \rightarrow N(0,\Sigma_0),
\end{equation}

\end{assumption}

and there exists a consistent estimate $\hat{\Sigma}_n$ for $\Sigma_0$.

\begin{assumption}
Let $\hat{\phi}_n$ be an estimate of $\phi_0$, where $\phi_0$ satisfies  
\begin{equation}
P |\hat{\phi}_n(A,X)-\phi_0(A,X)|^2\xrightarrow{P}0,\end{equation} 
where $P$ means taking the expectation over $(X,A,Y)$ under the true model.
\end{assumption}

In general, most generalize linear models satisfy all assumptions under regularity conditions. The assumptions can be verified using a Taylor expansion, the asymptotic normality of the maximum likelihood estimator, and standard empirical process arguments. For illustration, detailed derivations for the logistic regression case are provided in Appendix A. Note that in randomized trial data, the propensity score will be fixed for binary treatment $\pi(a|x)=0.5$ and $\hat{\theta}_n=0$.

\subsection{Value Function}

We evaluate the goodness of a decision rule $\hat{d}_{j,n}$ by estimating the value function $V(\hat{d}_{j,n})  = E(Y|A=\hat{d}_{j,n}(X),X)=E(\frac{Y1\{A=\hat{d}_{j,n}(X)\}}{P(A|X)})$ \cite{qian2011performance}. The value function can be interpreted as the inverse propensity score weighted average outcome if the population were to follow the decision rule $\hat{d}_{j,n}$.
Specifically, we will calculate the value function based on the test set data, represented by the subscript m:
\begin{equation*}
\hat{V}_{m}(\hat{d}_{j,n}(X_i))=\frac{\sum_{i=1}^m \frac{y_i 1\{A_i=\hat{d}_{j,n}(x_i)\}}{\hat{\pi}_n(A_i|X_i)}}{\sum_{i=1}^m\frac{1\{A_i=\hat{d}_{j,n}(x_i)\}}{\hat{\pi}_n(A_i|X_i)}},
\end{equation*}
where the propensity score $\hat{\pi}_n$ and the ITR $\hat{d}_{j,n}$ are estimated from the training set and applied to the test set.  
The data $(X_i,A_i,Y_i)$, $i=1,\dots,m$ come from the test set.  It is crucial to ensure independence between training and test sets for the test set results to effectively reflect the generalizability of the training set result. 

\subsection{Variance for the Value Function}

Let $Z=(X,A,Y)$. For any measurable function of $Z$, let $U = U(Z)$, $\mathbb{P}_{m}(U)=m^{-1}\sum_{i=1}^{m}U_i$ be the empirical measure, and $P(U)=E(U)$ be the expectation taken over $U$. Then
$\mathbb{G}_{m}(U) = \sqrt{m}[\mathbb{P}_{m}(U)-P(U)]\rightarrow N(0,Var(U))$ by standard empirical process arguments. Through some derivations we can derive the following equation (detailed proofs are attached in Appendix A):
\begin{equation}
\label{value_derivation}
    \begin{split}
    \sqrt{m}(\hat{V}_{m}(\hat{d}_{j,n})-V_0(\hat{d}_{j,n})) & = \mathbb{G}_{m}(\frac{(Y- V_0(\hat{d}_{j,n}))1\{A=\hat{d}_{j,n}(X)\}}{\pi_0(A|X)}) \\
    & -\sqrt{m/n}\sqrt{n}(\hat{\theta}_n-\theta_0)^T
    E(\phi_0(A,X)\frac{(Y-V_0(\hat{d}_{j,n}))1\{A=\hat{d}_{j,n}(X)\}}{\pi^2_0(A|X)})+o_P(1),\\
    \end{split}
\end{equation}
where m denotes the empirical estimate based on the test set with $m$ individuals.\\

Let $U=U(Z)=\frac{(Y-V_0(\hat{d}_{j,n}))1\{A=\hat{d}_{j,n}(X)\}}{\pi_0(A|X)}$ and $W = W(Z) = E(\phi_0(A,X)\frac{(Y-V_0(\hat{d}_{j,n}))1\{A=\hat{d}_{j,n}(X)\}}{\pi^2_0(A|X)})$. We can show that the following proposition holds.\\

\begin{proposition} \label{prop:vvar}
\begin{equation}
    \sqrt{m}(\hat{V}_{m}(\hat{d}_{j,n})-V_0(\hat{d}_{j,n})) \rightarrow N(0,Var(U)+\frac{m}{n} W^T \Sigma_0 W),
\end{equation}
provided that the quotient $m/n$ asymptotically approaches a finite limit.
\end{proposition}

Standard empirical process arguments yield $\mathbb{G}_{m}(U) = \sqrt{m}[\mathbb{P}_{n}(U)-P(U)] \rightarrow N(0,Var(U))$. Since $\sqrt{n}(\hat{\theta}_n-\theta_0) \rightarrow N(0,\Sigma_0)$, we can also show that the random variables $\sqrt{m}(\hat{V}_{m}(\hat{d}_{j,n})-V_0(\hat{d}_{j,n}))$ converge in distribution to a mean zero normal distribution. In this context, as the value of $n$ approaches infinity, we assume the quotient $m/n$ asymptotically approaches a finite limit, rather than diverging towards infinity. 

where the estimated influence function for $U$ and $W$ are as follows:

\begin{align}
 &\hat{U}_{i,j}  =(Y_{i}-\hat{V}_{n}(\hat{d}_{j,n}))\frac{1\{A_{i}=\hat{d}_{j,n}\}}{\hat{\pi}_{n}(A_i|X_i)},\\
 &\bar{U}_{j} = m^{-1}\sum_{i=1}^{m}\hat{U}_{i,j},\\
 &\hat{W}_{j}  = m^{-1}\sum_{i=1}^{m}[\frac{\hat{\phi}(A_i,X_i)\hat{U}_{i,j}}{\hat{\pi}_{n}(A_i|X_i)}], 
\end{align}

where $i = 1,...,n$.\\

Thus, the variance $\sigma_0^2$ for the value function $V_m(d_j)$ can be estimated consistently by:

\begin{equation}\label{eq:estvar}
\hat{\sigma}^2_{m,j}=Var(\hat{V}_{m}(\hat{d}_{j})) \approx m^{-2}\sum_{i=1}^{m}(\hat{U}_{i,j}-\bar{U}_{j})^2+\frac{1}{n}\hat{W}_{j}^T\hat{\Sigma}_{n}\hat{W}_{j}.
\end{equation}

Note that the two parts in the variance equation are dependent. When the number of individuals in the test set is much less than that in the training set ($m<<n$), then we can ignore the second term $\frac{1}{n}\hat{W}_{j}^T\hat{\Sigma}_{n}\hat{W}_{j}$.\footnote{When the number of individuals in the test set is much less than that in the training set, $m<<n$. Thus $m^{-1}>>n^{-1}$. Looking at equation (9),
$m^{-2}\sum_{i=1}^{m}(\hat{U}_{i,j}-\Bar{U}_{j})^2 >> \frac{1}{n}\hat{W}_{j}^T\hat{\Sigma}_{n}\hat{W}_{j}$.
} When we use randomized trial data, we don't need to estimate the propensity score, and thus $\hat{\Sigma}_n=0$ in this situation, and hence only the first term remains.

\subsection{Comparison Between Value Functions}
The distribution of the random variable for the difference between the two value functions of two ITRs $\hat{d}_{1,n}$ and $\hat{d}_{2,n}$ can be shown to converge to a normal distribution:

\begin{proposition}\label{prop:vvardiff}
\begin{equation}
    \sqrt{m}(\hat{V}_{m}(\hat{d}_{1,n})-\hat{V}_{m}(\hat{d}_{2,n})-V_0(\hat{d}_{1,n})+V_0(\hat{d}_{2,n})) \rightarrow N(0,T_{0}^2),
\end{equation}
where

\begin{equation*}
    T_{0}^2=Var(U_1-U_2)
    +\frac{m}{n}(W_{1}-W_{2})^T \Sigma_{0}(W_{1}-W_{2}).
\end{equation*}
\end{proposition}

Thus, the variance for the difference between the value function of two ITRs is approximately:

\begin{equation}\label{eq:estvardiff}
    Var(\hat{V}_{m}(\hat{d}_{1,n})-\hat{V}_{m}(\hat{d}_{2,n}))=m^{-2}\sum_{i=1}^{m}(\hat{U}_{i,1}-\hat{U}_{i,2}-\Bar{U}_{1}+\Bar{U}_{2})^2+\frac{1}{n}(\hat{W}_{1}-\hat{W}_{2})^T\hat{\Sigma}_{n}(\hat{W}_{1}-\hat{W}_{2}).
\end{equation}

Here, $\hat{\Sigma}_{n}$ is estimated from the training set based on both variances due to multiple imputation and variance due to modeling the sample data:
\begin{equation*}
    \hat{\Sigma}_n = nVar(\hat{\theta}_n) = nK^{-1}\sum_{k=1}^K Var(\hat{\theta}_{n,k})+n(1+1/K)(K-1)^{-1}\sum_{k=1}^K(\hat{\theta}_{n,k}-\bar{\theta}_{n})^2,
\end{equation*}
where $\Bar{\theta}_{n}=1/K \sum_{k=1}^K \hat{\theta}_{n,k}$. $\hat{\Sigma}_{n}$ is the same across imputations, and it is positive definite because it is the sum of a positive definite matrix and a positive semi-definite matrix. 
Under the null hypothesis, $V_0(\hat{d}_{1,n})-V_0(\hat{d}_{2,n})=0$. The t-test statistics $t$ for this null hypothesis can then be constructed as follows:
\begin{equation}\label{eq:ttest}
t = \frac{\hat{V}_{m}(\hat{d}_{1,n})-\hat{V}_{m}(\hat{d}_{2,n})}{\sqrt{Var(\hat{V}_m(\hat{d}_{1,n})-\hat{V}_m(\hat{d}_{2,n}))}} \sim N(0,1),
\end{equation}
and this can be used to compute the p-value.

\subsection{Multiple Imputation}
In the presence of missing data, we extend our method to address the variance of the value function in the case of multiple imputation. This extension is based on the ideas from Chapter 2.3.2 from Van Buuren (2018) \cite{van2018flexible}.
Suppose we have K imputations. Let $k=1,\dots,K$ be the index of the multiple-imputed data. The subscript m indicates that the estimates are obtained for the test set.

For the single decision rule $\hat{d}_{j}$, let $\hat{V}_{m,k}(\hat{d}_{j})$ denote the estimated value function for the $k^{th}$ imputed test set. Let $\Tilde{V}_{m}(\hat{d}_j)=K^{-1}\sum_{k=1}^K\hat{V}_{m,k}(\hat{d}_j)$, be value function estimate for decision rule $\hat{d}_j$. Let $\hat{\sigma}_{m,j,k}=Var(\hat{V}_{m,k}(\hat{d}_{j}))$ be the estimator of the variance-covariance matrix for the estimated value function for the $k^{th}$ imputed test set.  Let $\sigma_{0,K}^2 = Var_K(\hat{V}(\hat{d}_j))$ denote the variance associated with the estimate for a single value function. The empirical estimate of $\sigma_{0,K}^2$ is:
\begin{equation}
    \hat{\sigma}_{m,K}^2 = (1+1/K)(K-1)^{-1}\sum_{k=1}^K(\hat{V}_{m,k}(\hat{d}_j)-\Tilde{V}_m(\hat{d}_j))^2+K^{-1}\sum_{k=1}^K \hat{\sigma}^2_{m,j,k}.
\end{equation}
For the paired test between decision rules $\hat{d}_{1}$ and $\hat{d}_{2}$, let $\hat{T}^2_{m,k}=Var(\hat{V}_{m,k}(\hat{d}_{1})-\hat{V}_{m,k}(\hat{d}_{2}))$ denote the estimated variance for the difference between two value functions for the $k^{th}$ imputed test set. Let $T_{0,K}^2 = Var_K(\hat{V}(\hat{d}_1)-\hat{V}(\hat{d}_2))$ denote the variance for the difference between value functions for decision rule $\hat{d}_1$ and $\hat{d}_2$. The empirical estimate of $T_{0,K}^2$ is:

\begin{equation}\label{eq:estvardiffmi}
    \hat{T}_{m,K}^2 = (1+1/K)(K-1)^{-1}\sum_{k=1}^{K}(\hat{V}_{m,k}(\hat{d}_1)-\hat{V}_{m,k}(\hat{d}_2)-\Tilde{V}_{m}(\hat{d}_1)+\Tilde{V}_{m}(\hat{d}_2))^2+K^{-1}\sum_{k=1}^{K}\hat{T}^2_{m,k}.
\end{equation}
Under the null hypothesis, $H_0:V_{0,K}(\hat{d}_{1,n})-V_{0,K}(\hat{d}_{2,n})=0$. The t-test statistics $t_K$ for this null hypothesis can be constructed as follows:

\begin{equation}\label{eq:ttestmi}
t_K = \frac{K^{-1}\sum_{k=1}^K(\hat{V}_{m,k}(\hat{d}_{1,n})-\hat{V}_{m,k}(\hat{d}_{2,n}))}{\sqrt{Var_K(\hat{V}_{m,K}(\hat{d}_{1,n})-\hat{V}_{m,K}(\hat{d}_{2,n}))}} \sim N(0,1)
\end{equation}
The p-value for this t-test can be obtained from the test statistic $t_K$.

\section{Simulation}\label{sec:sim}
\begin{figure}[t]\centerline{\includegraphics[width=450pt,height=300pt]{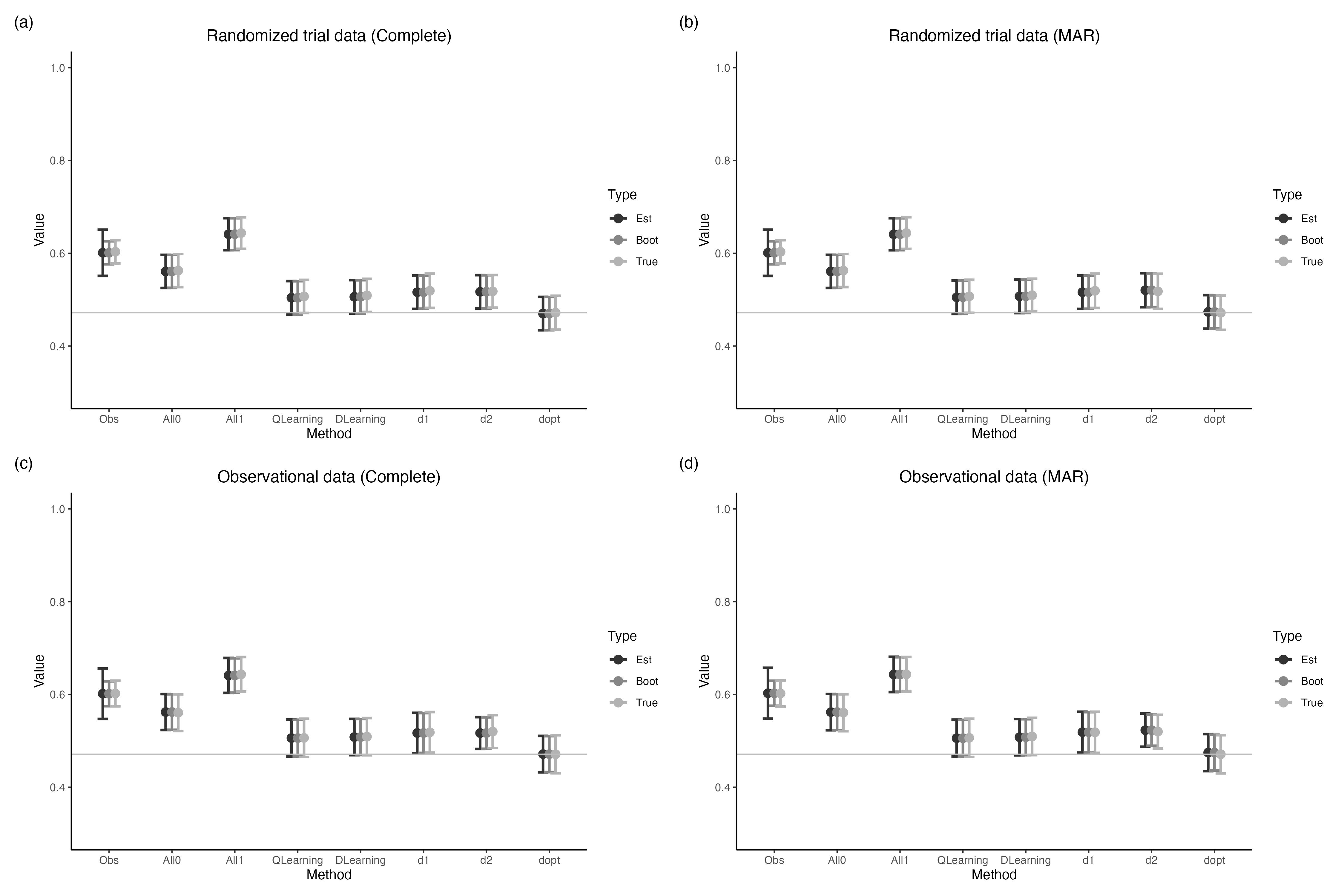}}
\caption{The value function results of four simulation scenarios with $\delta_t =1$ and $n = 5000$: (a) with random treatment assignment and complete data, (b) with random treatment assignment and data missing at random (c) with treatment assignment depending on the covariates and complete data, (d) with treatment assignment depending on the covariates and data missing at random. Within each scenario, we compare the value functions of eight different treatment regimes which are indicated in the x-axis: observed treatment, treatment $A=0$ for all individuals, treatment $A=1$ for all individuals, Q-learning optimal ITR, and D-learning optimal ITR, treatment rule $d_1$, treatment rule $d_2$, and the true optimal ITR. The true value function is presented with the grey line. The estimated values and 95\% confidence intervals from our method, bootstrap estimates, and the true empirical 95\% confidence intervals are depicted using varying shades of gray.}
\label{fig:sim_value}
\end{figure}

\begin{figure}[t]\centerline{\includegraphics[width=450pt,height=300pt]{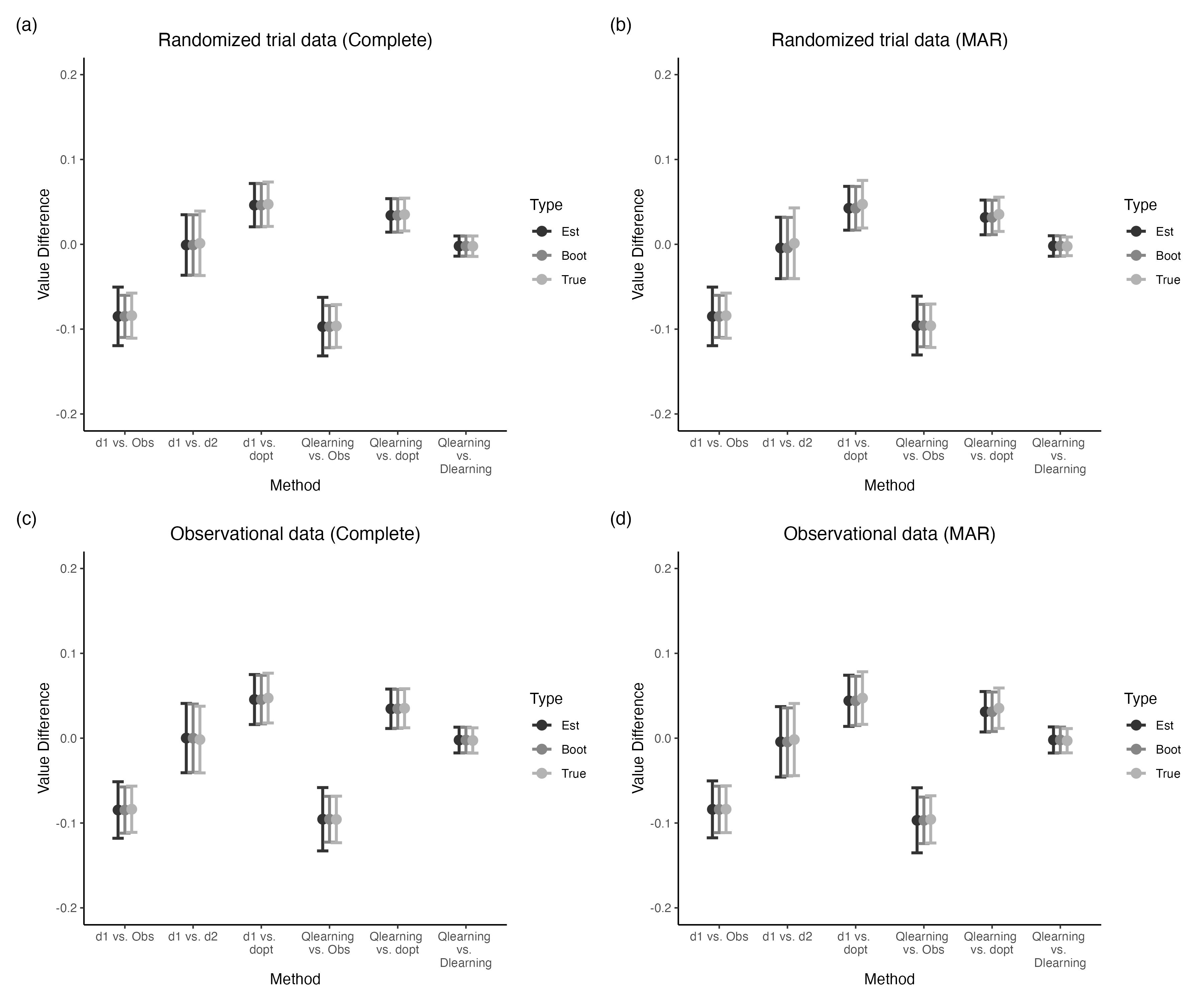}}
\caption{The differences of value functions and the associate standard deviation between selected pairs of the following treatments: observed treatment, universal treatment $A=0$, universal treatment $A=1$, Q-learning's optimal ITR, D-learning's optimal ITR, treatment rule $d_1$, treatment rule $d_2$, and the true optimal ITR. The results are presented under four simulation scenarios with $\delta_t = 1$ and $n = 5000$, comparing our estimates, bootstrap estimates, and the true values and 95\% confidence intervals.}
\label{fig:sim_valuediff}
\end{figure}

We evaluate and compare the value functions obtained from applying each of the following treatment regimes to the population under two scenarios: (1) the observed treatment, (2-3) the one-size-fits-all approach (one for each of the binary treatments), (4) Q-learning ITR, (5) D-learning ITR. A convex penalty, elastic net penalization \citep{zou2005regularization}, was used for each of the ITR models for variable selection. We generate a total of 5000 observations for each simulated training set and 1500 observations for each test set. The number of covariates included in the model is $p=20$. We perform 1000 replicates for each of the simulations.

The following data generation process is modified from the first simulation scenario in Tian \textit{et al} \cite{tian2014simple}. The covariates $X$ are generated from three different underlying distributions: $X_1,...,X_{10}$ are uncorrelated and generated from a multivariate normal distribution with mean zero and the covariance-variance matrix is a diagonal of 1s; $X_{11},...,X_{15}$ are generated from a uniform distribution $U(-0.5,0.5)$; $X_{16},...,X_{20}$ are generated from a binary distribution with $p=0.5$. Let $A$ denote the binary treatment. In the model implementation step, we will use $A=\{0,1\}$ for Q-learning and switch to $A=\{-1,1\}$ for D-learning. $\varepsilon \sim N(0,1)$ is the random error\footnote{The error term has been incorporated into the outcome data generation mechanism to better reflect potential real-world inaccuracies.} for generating the outcome $Y$. Let $\beta_0=(\sqrt{6})^{-1}$, $\beta_1=\beta_2=0$, $\beta_j=(2\sqrt{6})^{-1}$, $j=3,4,\dots,10$, $\beta_{11}=...=\beta_{p} =0$ be the coefficient for the main effect. Let $Y\sim binomial$ with probability equal to the expit of
\begin{equation*}
(\beta_0+\sum_{j=1}^p\beta_jX_j)^2+(I(X_1 > 0) + I(X_2 > 0) - 2I(X_1 \leq 0)I( X_2 \leq 0))A \delta_t+ \varepsilon.
\end{equation*}

The generated data is designed to have main effects $(\beta_0,\dots,\beta_p)$, and treatment interaction effect $(I(X_1 > 0) + I(X_2 > 0) - 2I(X_1 \leq 0)I( X_2 \leq 0))\delta_t$, where $\delta_t>0$ is a parameter that controls the magnitude of the treatment interaction effect on the outcome. To align with the application example presented in section \ref{sec:appli}, we assume, without loss of generality, that a smaller outcome Y and a smaller value function are preferable. Thus, the optimal treatment regime is $d_{opt} = I(I(X_1 > 0) + I(X_2 > 0) - 2I(X_1 \leq 0)I( X_2 \leq 0)) \leq 0=I(X_1 \leq 0\, and\, X_2 \leq 0)$. We consider four scenarios:

(a) Propensity score $\pi=0.5$ (randomized trial data) 

(b) Propensity score $\pi=0.5$ (randomized trial data with data missing at random)

(c) Propensity score $logit(\pi) = 0.75X_1-0.75X_2$ (observational data)

(d) Propensity score $logit(\pi) = 0.75X_1-0.75X_2$ (observational data with data missing at random)

In scenarios where data are missing at random, the missingness mechanism is implemented by assigning a conditional probability of missingness to the variables $X_2, \ldots, X_p$. This probability is contingent upon the value of $X_1$: specifically, there is a 15\% probability of missingness when $X_1 > 0$, and a lower probability of 10\% in cases where this condition is not met.

The true value function $V_*(\hat{d}_{j})$ is obtained from applying each decision rule $\hat{d}_{j}$ on a simulated large test set comprising 100,000 individuals. The data generation process for this large test set mirrors that of the corresponding complete training set. The true value function $V_*(\hat{d}_{j})$ is calculated based on the true propensity score parameters and the decision rule $\hat{d}_{j}$ derived from the training set. We define the true variance of the value function as $\hat{\tau}_{j}^2 = \frac{1}{L} \sum_{l=1}^{L} (V_{l} (\hat{d}_{j}) -V_*(\hat{d}_{j}))^2$, where  $V_{l} (\hat{d}_{j}) $ represents the estimated value function from test set $l$, utilizing the estimated decision rule $\hat{d}_{j}$ and the propensity score model from the corresponding training set. The true variance for the difference between value functions is constructed in a similar manner. 

The point estimates for value functions are derived by averaging the value function estimates across 1000 replications on the test set. The estimated variance for each value function is computed using Equation \ref{eq:estvar}, while the variances for differences between two value functions are calculated using Equations \ref{eq:estvardiff} (for scenarios 1 and 3) and \ref{eq:estvardiffmi} (for scenarios 2 and 4), also averaged over 1000 replications. For bootstrapping estimates, the point estimate of the value function is the mean obtained from 1000 bootstrap samples (we have also obtained estimates from 100, 200, 500 bootstrapping samples, the results are similar to 1000 bootstraping samples and are thus omited here), which is then further averaged across 1000 replications. Similarly, the bootstrapping variance is determined by calculating the variance of the 1000 bootstrap samples and averaging this across 1000 replications.

Figure \ref{fig:sim_value} shows the value functions and standard deviation in four scenarios (a)-(d). We observe that the point estimates of value functions using our estimated method are close to that estimated from bootstrapping and the true value, across all 4 scenarios and almost all methods. The variance estimates of a single value function using our estimated method are close to that estimated from bootstrapping and the true value (except that our method gives a slightly larger variance estimate for the observed treatment). Figure \ref{fig:sim_valuediff} shows that, for comparing pairwise value functions, the point estimate and the variance estimated for the difference from our method are also close to the bootstrapping estimates and the truth. As expected, the value function for ITRs from $d_{opt}$ is the lowest among all methods. Table \ref{tab:sim_test_diff} presents the difference, standard deviation (SD), and t-test rejection rates for comparisons between $d_1$ vs. $d_2$ and $d_1$ vs. $d_{opt}$, under the setting $\delta_t = 1$ and $n=5000$. The point estimates and SDs are consistent across our estimates, bootstrapped estimates, and the true value differences. The t-test for the difference between $d_1$ and $d_2$ yields a rejection rate close to zero, whereas the test for the difference between $d_1$  and $d_{opt}$ results in a rejection rate close to one. For observational data, when treatment assignments are nonrandom and probabilities need to be estimated by propensity scores, value function estimates have slightly larger variances compared to the value function estimates for randomized trial data (Table \ref{tab:sim_test_diff}). For datasets with missing values, the performance remains comparable to that of complete datasets when the missing data is random and is addressed through multiple imputation using missForest. The results shown here are the test set results. The training set results are similar to the test set results but with a smaller variance. The simulation results illustrate the effectiveness of our approach in discerning disparities between two ITRs while maintaining reasonable variance estimates in both observational data and randomized trial data, with or without missing values.

\begin{center}
\begin{table*}[t]%
\caption{Estimated difference in value functions Mean (SD) and t-test rejection rate (RR) under four scenarios for simulation data based on 1000 replications ($\delta_t = 1$, $n=5000$).\label{tab:sim_test_diff}}
\centering
\resizebox{\textwidth}{!}{%
\begin{tabular}{llcccccccc}
\toprule
&& \multicolumn{2}{c}{\textbf{Scenario 1}} & \multicolumn{2}{c}{\textbf{Scenario 2}} & \multicolumn{2}{c}{\textbf{Scenario 3}} & \multicolumn{2}{c}{\textbf{Scenario 4}}\\
\cmidrule{3-4} \cmidrule{5-6} \cmidrule{7-8} \cmidrule{9-10}
\textbf{Method} & \textbf{Test} & \textbf{Mean(SD)} & \textbf{RR} & \textbf{Mean(SD)} & \textbf{RR} & \textbf{Mean(SD)} & \textbf{RR} & \textbf{Mean(SD)} & \textbf{RR} \\

\midrule
Est & $d_1$ vs. $d_2$ & -0.001(0.018) & 0.069 & -0.004(0.018) & 0.089 & 0.000(0.021) & 0.039 & -0.004(0.021) & 0.055\\
Boot & $d_1$ vs. $d_2$ & -0.001(0.018) & 0.069 & -0.004(0.018) & 0.085 & 0.000(0.021) & 0.041 & -0.004(0.020) & 0.062\\
True & $d_1$ vs. $d_2$ & 0.001(0.019) & 0.000 & 0.001(0.021) & 0.000 & -0.002(0.020) & 0.000 & -0.002(0.022) & 0.000\\
\addlinespace
Est & $d_1$ vs. 
$d_{opt}$ & 0.046(0.013) & 0.940 & 0.043(0.013) & 0.896 & 0.046(0.015) & 0.861 & 0.044(0.015) & 0.808\\
Boot & $d_1$ vs. 
$d_{opt}$ & 0.046(0.013) & 0.939 & 0.043(0.013) & 0.900 & 0.045(0.015) & 0.869 & 0.044(0.015) & 0.822\\
True & $d_1$ vs. 
$d_{opt}$ & 0.047(0.013) & 1.000 & 0.047(0.014) & 1.000 & 0.047(0.015) & 1.000 & 0.047(0.016) & 1.000\\
\bottomrule
\end{tabular}
}
\end{table*}
\end{center}

For our power analysis, we define two regimes: $d_1 = I(X_1 \leq 0)$, $d_2 = I(X_2 \leq 0)$. By design, $V_0(\hat{d}_{1,n}) = V_0(\hat{d}_{2,n})$ and $V_0(\hat{d}_{1,n}) \neq V_0(\hat{d}_{opt,n})$. The power analysis includes different training sample sizes $n \in\{ 1000, 2000, 3500,5000\}$ and varying values of $\delta_t \in \{1,2,5,10\}$, where treatment interaction hyperparameter $\delta_t$ controls the magnitude of treatment heterogeneity. As $\delta_t$ increases, the difference between value functions of $d_1$ and $d_{opt}$ also increases. 
Table \ref{tab:sim_type1_by_ntrain} show the Type I Error rate by testing $H_0: d_1 = d_2$. Table \ref{tab:sim_power_by_ntrain} shows the power (or 1-Type II Error rate) by testing $H_0: d_1 = d_{opt}$. These results show that both the Type I Errors and the power of our method are slightly lower or similar to that of bootstrapping method across 4 scenarios, across different effect size and different sample sizes.\footnote{The results for $\delta_t\in\{5,10\}$ were similar, we only present the results for $\delta_t\in\{1,2\}$ for clarity.}

\begin{center}
\begin{table}
\caption{Type 1 Error by sample size and $\delta_t$. \label{tab:sim_type1_by_ntrain}}
\centering
\begin{tabular}[t]{llrrrrrrrr}
\toprule
 & & \multicolumn{2}{@{}c@{}}{\textbf{Scenario 1}} & \multicolumn{2}{@{}c@{}}{\textbf{Scenario 2}} & \multicolumn{2}{@{}c@{}}{\textbf{Scenario 3}} & \multicolumn{2}{@{}c@{}}{\textbf{Scenario 4}}\\\cmidrule{3-4}\cmidrule{5-6} \cmidrule{7-8} \cmidrule{9-10}
\textbf{n} & \textbf{Method} & \textbf{$\delta_t$ = 1} & \textbf{$\delta_t$ = 2} & \textbf{$\delta_t$ = 1}  & \textbf{$\delta_t$ = 2} & \textbf{$\delta_t$ = 1}  & \textbf{$\delta_t$ = 2} & \textbf{$\delta_t$ = 1}  & \textbf{$\delta_t$ = 2}\\
\midrule
1000 & Est & 0.041 & 0.054 & 0.061 & 0.070 & 0.038 & 0.037 & 0.052 & 0.041\\
1000 & Boot & 0.047 & 0.060 & 0.066 & 0.075 & 0.060 & 0.075 & 0.077 & 0.086\\
\addlinespace
2000 & Est & 0.053 & 0.057 & 0.062 & 0.087 & 0.049 & 0.044 & 0.062 & 0.063\\
2000 & Boot & 0.058 & 0.063 & 0.069 & 0.090 & 0.059 & 0.066 & 0.079 & 0.084\\
\addlinespace
3500 & Est & 0.059 & 0.064 & 0.086 & 0.085 & 0.049 & 0.051 & 0.057 & 0.067\\
3500 & Boot & 0.062 & 0.064 & 0.085 & 0.085 & 0.056 & 0.057 & 0.067 & 0.076\\
\addlinespace
5000 & Est & 0.069 & 0.063 & 0.089 & 0.097 & 0.039 & 0.039 & 0.055 & 0.071\\
5000 & Boot & 0.069 & 0.065 & 0.085 & 0.097 & 0.041 & 0.047 & 0.062 & 0.076\\
\bottomrule
\end{tabular}
\end{table}
\end{center}

\begin{center}
    \begin{table}

\caption{Power by sample size and $\delta_t$. \label{tab:sim_power_by_ntrain}}
\centering
\begin{tabular}[t]{rlrrrrrrrr}
\toprule
 & & \multicolumn{2}{@{}c@{}}{\textbf{Scenario 1}} & \multicolumn{2}{@{}c@{}}{\textbf{Scenario 2}} & \multicolumn{2}{@{}c@{}}{\textbf{Scenario 3}} & \multicolumn{2}{@{}c@{}}{\textbf{Scenario 4}}\\\cmidrule{3-4}\cmidrule{5-6} \cmidrule{7-8} \cmidrule{9-10}
\textbf{n} & \textbf{Method} & \textbf{$\delta_t$ = 1} & \textbf{$\delta_t$ = 2} & \textbf{$\delta_t$ = 1}  & \textbf{$\delta_t$ = 2} & \textbf{$\delta_t$ = 1}  & \textbf{$\delta_t$ = 2} & \textbf{$\delta_t$ = 1}  & \textbf{$\delta_t$ = 2}\\
\midrule
1000 & Est & 0.916 & 1.000 & 0.850 & 0.997 & 0.811 & 0.998 & 0.757 & 0.996\\
1000 & Boot & 0.925 & 1.000 & 0.860 & 0.998 & 0.879 & 0.999 & 0.838 & 0.999\\
\addlinespace
2000 & Est & 0.925 & 0.999 & 0.870 & 0.999 & 0.876 & 1.000 & 0.833 & 1.000\\
2000 & Boot & 0.929 & 0.999 & 0.875 & 0.999 & 0.899 & 1.000 & 0.869 & 1.000\\
\addlinespace
3500 & Est & 0.918 & 1.000 & 0.864 & 1.000 & 0.880 & 0.999 & 0.846 & 0.998\\
3500 & Boot & 0.917 & 1.000 & 0.866 & 1.000 & 0.892 & 0.999 & 0.866 & 0.999\\
\addlinespace
5000 & Est & 0.940 & 1.000 & 0.896 & 1.000 & 0.861 & 1.000 & 0.808 & 1.000\\
5000 & Boot & 0.939 & 1.000 & 0.900 & 1.000 & 0.869 & 1.000 & 0.822 & 1.000\\
\bottomrule
\end{tabular}
\end{table}
\end{center}

Since our method relies on the correct specification of the propensity model, we examine how propensity score estimation influences the value function and standard deviation estimates (Figure \ref{fig:sim_value_propen}) as well as the differences between value functions and their associated standard deviation estimates (Figure \ref{fig:sim_valuediff_propen}). The sensitivity analysis indicates that the results remain similar whether using the estimated propensity score or the true propensity score. However, our method may exhibit bias when the propensity score is misspecified; however, the extent of this bias is comparable to that observed in the bootstrap estimates. These findings highlight a limitation of our approach: the accuracy of variance estimation depends on the precision of propensity score estimation. Additionally, they provide insights into how our method performs under model misspecification in certain conditions (Figure \ref{fig:propen_demo} illustrates the distribution of selected propensity score models in relation to the true propensity score across the these four models).


\section{Application}\label{sec:appli}
\subsection{CHNS: Data Analysis}

Hypertension, also known as high blood pressure, is a chronic medical condition that affects over one billion adults worldwide \citep{bloch2016worldwide}. It is a major risk factor for cardiovascular disease, which is the leading cause of death globally \citep{bromfield2013high}. People with hypertension are more likely to develop kidney disease, \citep{weldegiorgis2020ckd} vision problems, \citep{bhargava2012eye} and cognitive impairment. \citep{kilander1998htn} Hypertension can also affect quality of life and increase annual medical costs \citep{wang2017annual}. Over the past two to three decades, China has experienced a significant increase in the prevalence of hypertension (from 20.8\% in 2004 to 29.6\% in 2010 and 24.7\% in 2018) due to increased life expectancy and lifestyle changes \citep{zhang2023prevalence, wang2023hypertension}. As hypertension has become a significant burden on China’s population health, it is important to study this condition and develop effective prevention and treatment strategies. 
While medication proves effective in managing hypertension, the general adherence to anti-hypertensive medication remains low. This can be attributed to factors such as a low reimbursement ratio, high daily medical costs, and access to healthcare \cite{cui2020analysis}. In addition to taking medications, adopting non-pharmacologic interventions such as engaging in regular physical activity, reducing salt intake, and following a balanced diet rich in fruits and vegetables can help to reduce high blood pressure \citep{appel2003lifestyle}. However, there is potential heterogeneity in which individuals may see the most benefit from behavioral interventions.  

Our overall objective is to identify the most effective non-pharmacologic treatments for subpopulations of individuals, considering the challenges of adhering to a comprehensive healthy lifestyle based on public health recommendations. As a first step, we examine the potential variability in the effectiveness of exceeding 5 hours of weekly moderate-to-vigorous physical activity (MVPA) in reducing the risk of hypertension in the CHNS population. We seek to distinguish between participants who benefit from engaging in MVPA for more than 5 hours per week and those who do not. For participants who see no hypertension benefit from increasing their physical activity, it is possible to identify other potential interventions to consider, such as diet modification. Understanding which interventions confer optimal outcomes empowers individuals to identify the most effective and efficient behavioral treatments and their associated benefits. This knowledge can be highly beneficial, as individuals are more motivated to engage in behaviors when they are aware of their positive outcomes \cite{cane2012validation}. To address this research question, we tested whether a personalized physical activity intervention strategy could achieve population-level health outcomes comparable to an intervention that assigns high physical activity to everyone.

We used data from the China Health and Nutrition Survey (CHNS)\cite{zhang2014c}, a population-based, observational data set consisting of high-quality data on diet (3-day repeated 24-hour recalls) and physical activity (detailed 7-day recall)  collected from individuals in China using detailed recall instruments. Our analysis focused on the study year 2009 with 8320 adults. Our outcome, hypertension, was defined \footnote{Anti-hypertensive medication is included in the model in our analysis, so blood pressure measurements here are not adjusted for medication. Participants who were taking medication may exhibit varying responses to changes in physical activity.  Consequently, we have included medication as a covariate that has the potential to influence how individuals respond to engaging in more than 5 hours of physical activity per week.} as a systolic blood pressure $\geq$ 130 mmHg or diastolic blood pressure $\geq$ 80 mmHg. Individuals with missing systolic or diastolic blood pressure measurements or who were missing more than 10\% of their covariates were excluded from our study, resulting in a final analytic sample of 5241 individuals, 60\% of whom had hypertension. In the CHNS dataset, physical activity encompassing leisure, occupational, transportation, and domestic activities, were collected via a comprehensive survey recall instrument \cite{ng2014physical}. Our study focused on weekly moderate-to-vigorous physical activity (MVPA), defined as activities with a minimum of 3 METS (Metabolic equivalent of task) \cite{ainsworth2011} \footnote{The METS, or Metabolic equivalent of task, is a unit of measurement for physical activity. One MET is equivalent to a person's oxygen consumption at a rate of 3.5 milliliters per kilogram per minute. \cite{sylvia2014practical}}. Our sample population has a high MVPA level, with a median of 15 hours per week \citep{ng2014physical}. Targeting the 5-hour MVPA weekly guideline \citep{PAguideline, piercy2018physical,ng2014physical}, we dichotomized the treatment into a binary variable $A$: over 5 hours ($A=1$) and 5 or fewer hours ($A=0$) of MVPA per week. In the CHNS data, 66\% of the sample population engaged in more than 5 hours of MVPA weekly. With insights from biological understanding and published literature \cite{ng2014physical}, we identified twenty risk factors for hypertension. These factors also have the potential to influence the relationship between physical activity and hypertension, including: age, gender, caloric intake, BMI, education, smoking status, sodium intake, potassium intake, alcohol consumption, hypertension medication, household income, province, and urbanization index. Less than 15\% of our sample had missing covariates, handled using missForest package in R for multiple imputation \citep{missForestManual, missForestPaper}. Table \ref{tab1} provides further details about these factors based on one imputed dataset. The other imputed datasets exhibit similar distributions. The data was split into a 70\% training set and a 30\% test set, with missForest imputation performed 10 times on each set to avoid dependency. We obtained Q-learning optimal ITRs \citep{qian2011performance} and D-learnings optimal ITR \citep{tian2014simple} for each of the training sets using a logistic regression model with elastic-net penalization. We then compared the value function of each treatment rule on the test set, including the observed treatment, assigning all individuals to treatment $A=1$ or $A=0$, optimal ITRs derived from Q-learning and D-learning.

\begin{center}
\begin{table}[t]%
\centering
\caption{Descriptive Table for CHNS Data.\label{tab1}}%
\resizebox{\textwidth}{!}{%
\begin{tabular}{lccc}
\toprule
\textbf{~} & \textbf{Training Data}  & \textbf{Test Data}  & \textbf{Overall}  \\
~ & (N=3667) & (N=1574) & (N=5241) \\
\midrule
Age, mean (SD) & 48.7 (10.4) & 48.8 (10.2) & 48.7 (10.3)  \\
Gender, n (\%) & ~ & ~ & ~ \\ 
\quad Male & 1695 (46.2\%) & 746 (47.4\%) & 2441 (46.6\%) \\ 
\quad Female & 1972 (53.8\%) & 828 (52.6\%) & 2800 (53.4\%) \\ 
Calorie Intake (kcal/kg), mean (SD) & 36.7 (11.8) & 36.7 (11.8) & 36.7 (11.8) \\ 
BMI, mean (SD) & 23.3 (3.18) & 23.4 (3.31) & 23.3 (3.22) \\ 
Education, mean (SD) & 1.56 (0.710) & 1.54 (0.723) & 1.56 (0.714) \\
Log-transformed Household Income, mean (SD) & 2.51 (0.853) & 2.50 (0.849) & 2.51 (0.852) \\ 
Current Smoking, n (\%) & 1234 (33.7\%) & 510 (32.4\%) & 1744 (33.3\%) \\ 
Sodium (mg/day), mean (SD) & 4560 (2270) & 4600 (2250) & 4570 (2260) \\ 
Potassium (mg/day), mean (SD) & 1750 (673) & 1720 (657) & 1740 (668) \\ 
Alcohol, n (\%) & 1262 (34.4\%) & 572 (36.3\%) & 1834 (35.0\%) \\ 
MVPA (hours/week) \tnote{$\dagger$}, mean (SD) & 34.2 (42.0) & 36.3 (43.3) & 34.8 (42.4) \\ 
Anti-Hypertensive Medication, n (\%) & 291 (7.9\%) & 144 (9.1\%) & 435 (8.3\%)\\
Hypertension, n (\%) & 2184 (59.6\%) & 940 (59.7\%) & 3124 (59.6\%) \\ 
Province, n (\%) & ~ & ~ & ~ \\ 
\quad 21 – Liaoning & 400 (10.9\%) & 170 (10.8\%) & 570 (10.9\%) \\ 
\quad 23 – Heilongjiang & 464 (12.7\%) & 199 (12.6\%) & 663 (12.7\%) \\ 
\quad 32 – Jiangsu & 424 (11.6\%) & 159 (10.1\%) & 583 (11.1\%) \\ 
\quad 37 – Shandong & 422 (11.5\%) & 200 (12.7\%) & 622 (11.9\%) \\ 
\quad 41 – Henan & 335 (9.1\%) & 148 (9.4\%) & 483 (9.2\%) \\
\quad 42 – Hubei & 367 (10.0\%) & 172 (10.9\%) & 539 (10.3\%) \\ 
\quad 43 – Hunan & 444 (12.1\%) & 183 (11.6\%) & 627 (12.0\%) \\ 
\quad 45 – Guangxi & 415 (11.3\%) & 156 (9.9\%) & 571 (10.9\%) \\ 
\quad 52 – Guizhou & 396 (10.8\%) & 187 (11.9\%) & 583 (11.1\%) \\
Urbanization Index, mean (SD) & 66.1 (19.0) & 65.3 (18.7) & 65.9 (18.9) \\
\bottomrule
\end{tabular}
}
\begin{tablenotes}
\item Source: China Health and Nutrition Survey (CHNS).
\item[$\dagger$]  MVPA includes occupational, domestic, transportation and leisure activity.

\end{tablenotes}
\end{table}
\end{center}

\begin{figure}[t]
\centerline{\includegraphics[width=450pt,height=300pt]{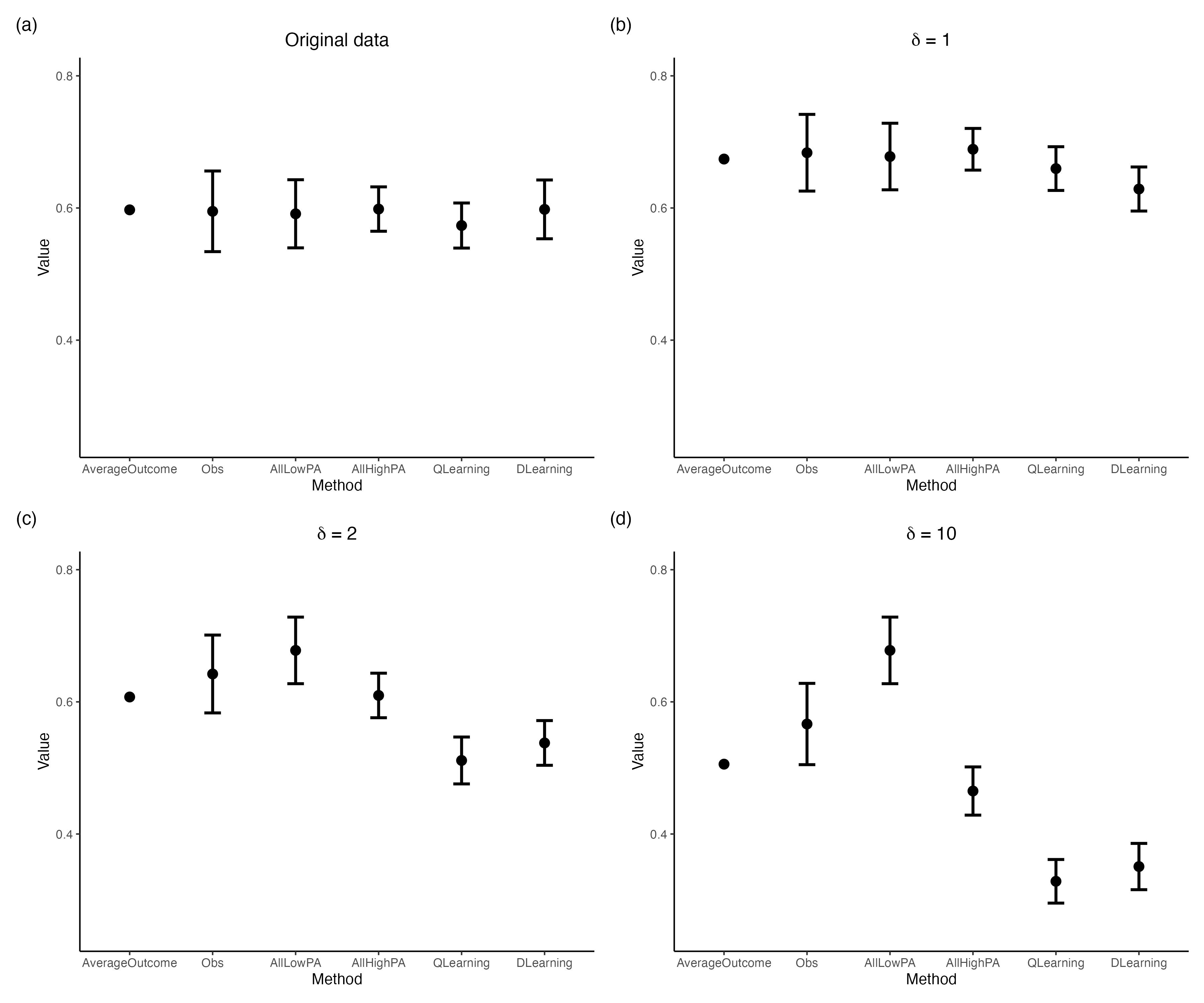}}
\caption{The value function and 95\% confidence interval estimates of six different treatment regimes: observed treatment (Obs), lower physical activity hours (AllLowPA, $A=0$) for all individuals, higher physical activity hours (AllHighPA, $A=1$) for all individuals, Q-learning optimal ITR (Q), and D-learning optimal ITR (D). For comparison purposes, we include the average outcome as a benchmark. (a) shows the results on original data, and (b)(c)(d) shows the three simulated data with three levels of treatment effect modification $\delta=1$, $\delta=2$, $\delta=10$, respectively. \label{fig:chns_value}}
\end{figure}

\begin{figure}[t]\centerline{\includegraphics[width=450pt,height=300pt]{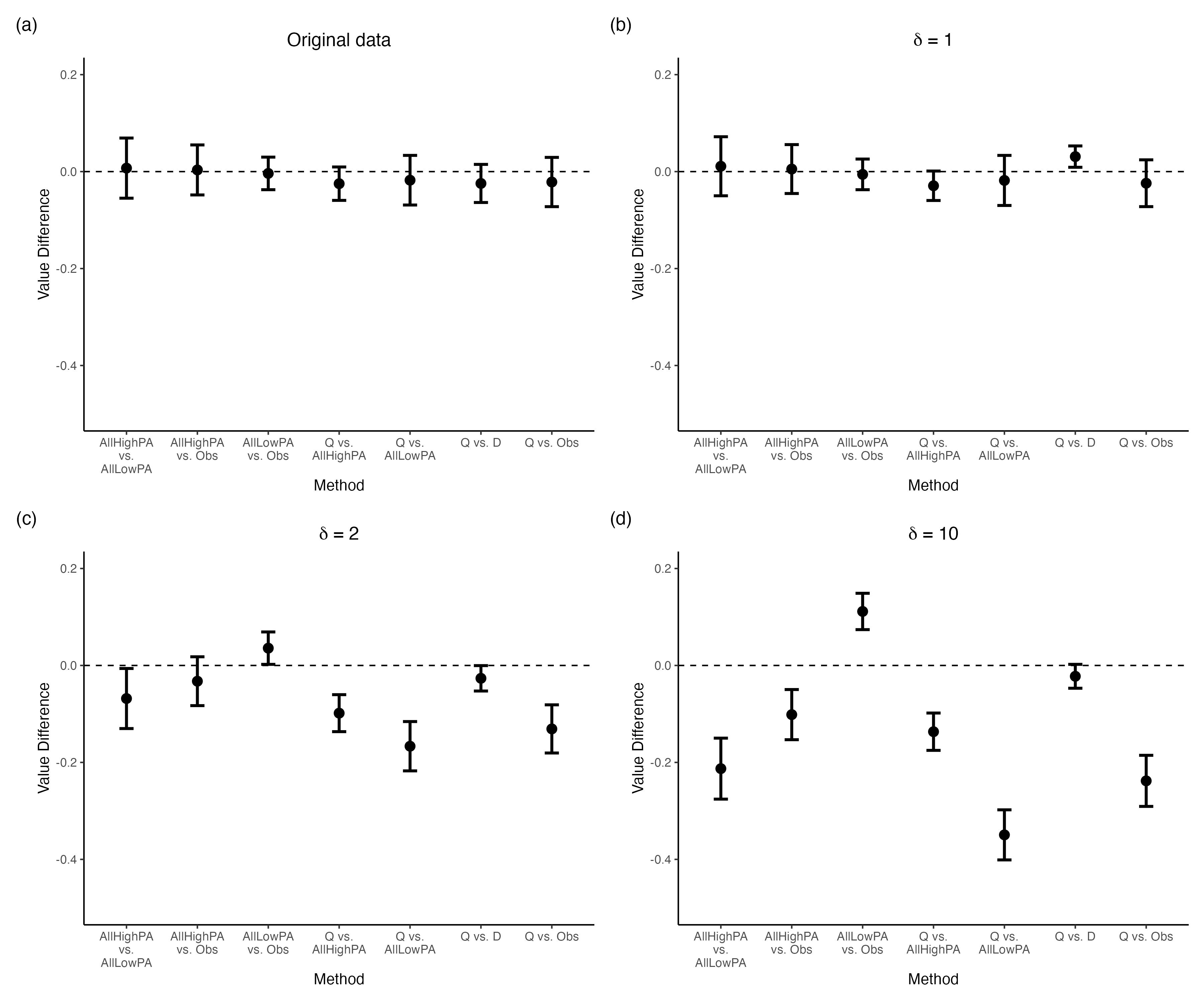}}
\caption{The value function differences and 95\% confidence interval among selected pairs of treatment regimes. Specifically, the Q-learning optimal ITR comparison with the observed treatment, treatment $A=0$ for all individuals, treatment $A=1$ for all individuals, and D-learning optimal ITR are shown. (a) shows the results on original data, and (b)(c)(d) shows the three simulated data with three levels of treatment effect modification $\delta=1$, $\delta=2$, $\delta=10$, respectively.}
\label{fig:chns_valuediff}
\end{figure}
 The value functions of these treatment rules were estimated using the common estimator $V$ and the differences between value functions were directly obtained by subtracting the value functions and then averaged over 10 imputed data sets. The variance of the value function and the variance of the difference between the value functions were estimated by our method, which accounts for both modeling and multiple imputation variance.  

Our goal is to compare personalized and uniform interventions in reducing hypertension risk by comparing the value functions in test sets for (1) observed physical activity, under the assumption that individuals continue their existing routines; (2) assigning all individuals to treatment $A=1$, assuming all individuals are above the 5 hours weekly MVPA recommendation; (3) assigning all individuals to treatment $A=0$, assuming all individuals are at or below the 5 hours weekly MVPA recommendation; and (4) optimal ITRs derived from various methods. Figure \ref{fig:chns_value}(a) shows the value function and 95\% confidence interval for those treatment regimes. The differences between the value functions for different decision rules are small. Especially, the Q-learning value function is the lowest but the confidence interval intersects with the that of other treatment regimes. This means that if the population follows the Q-learning ITR, the expected risk of hypertension in the population is similar to when the population follows the other treatment assignment (e.g. high PA for everyone). Figure \ref{fig:chns_valuediff}(a) shows the pairwise difference of the value functions between treatment regimes that we are interested in. Our focus is on determining the necessity of recommending over 5 hours of weekly MVPA to everyone or specific subgroups. Thus we compare the best ITR, Q-learning ITR, defined by the lowest value function estimate in the training set (Table \ref{tab:chns_original_value}) with the one-size-fits-all treatment $A=1$ for the test set. Based on the t-test result, the value functions are not significantly different, with a p-value of 0.156 for the Q-learning personalized treatment rule compared to assigning all individuals to treatment $A=1$ (Table \ref{tab:chns_original_valuediff}). This suggests that a personalized approach can achieve a similar hypertension risk as the population-level physical activity recommendations for doing physical activity for more than five hours per week.The results presented herein pertain to the test set, which is essential for evaluating the model’s generalizability to new, unseen data. A comparison of model performance between the training and test sets is provided in Appendix Figure \ref{fig:chns_valuesupp} and Figure \ref{fig:chns_valuediffsupp}. The training set results closely mirror those of the test set, underscoring the model’s stability; however, the test set results are emphasized as they more accurately reflect the model’s anticipated performance in real-world applications.

Our findings indicate that applying a uniform high level of physical activity across the entire population does not significantly reduce hypertension risk compared to tailoring activity levels to individual characteristics. Specifically, our results suggest that increased physical activity may benefit certain subgroups, whereas others do not derive additional benefits beyond 5 hours of weekly MVPA. For these subpopulations, alternative non-pharmacologic interventions may be more effective in mitigating hypertension risk.  For instance, individuals with high sodium consumption may not benefit significantly from increased physical activity alone. In contrast, sodium intake reduction could potentially offer a more effective strategy for lowering hypertension risk. Another possible explanation is that much physical activity in the CHNS data comprised working, as opposed to leisure, activities. Published data suggest that occupational physical activity may not be as protective against hypertension \cite{huai2013physical}. The complete pairwise comparison of all the comparisons is depicted in the appendix Figure \ref{fig:chns_valuediffsupp}.

\subsection{CHNS: Augmented Simulation Analysis}

In Section 4.1, our method did not detect a significant difference between the two treatment rules under comparison. This finding prompted the question of whether the null result was attributable to limitations of the method when applied to real-world data, or if the true difference was indeed minimal. To address this issue, we conducted an augmented analysis to evaluate our method's performance in scenarios where a true difference between the treatment rules is present. It is important to note that the models in Section 4.1 considered only a limited set of factors influencing the effect of physical activity on hypertension. Given the complexity of hypertension—arising from the interplay of individual behavior, environmental influences, and additional factors such as metabolism and genetics—incorporating these variables could reveal important insights and better identify the underlying heterogeneity in the impact of physical activity on hypertension. Consequently, we examined our method’s performance in the presence of prescriptive treatment effect interactions that account for these additional factors, or their combinations with our existing factors, which may explain more of the variation in individual hypertension risk. An additional analysis was therefore conducted using simulated datasets with modified prescriptive treatment effect interaction terms.  

 The outcome variables in these data sets were generated by the fitted Q-learning outcome regression model with the treatment effect interaction terms (excluding the treatment main effect term) multiplied by a factor of $\delta \geq 1$. We investigate three options of $\delta$ for this analysis: $\delta=1$, $\delta=2$, and $\delta=10$, representing the estimated original, double, and tenfold treatment effect, respectively. In each analysis, hypertension outcomes are replaced by generated outcomes from the modified Q-learning outcome regression model. This augmented analysis examines our method's performance when the outcome generative process is known and how the performance varies as the true prescriptive treatment effect enlarges. As $\delta$ increases, from $\delta=1$, to $\delta=2$, to  $\delta=10$ (Figure \ref{fig:chns_value} (b)-(d)),  the value function differences between the Q-learning strategy and other treatment rules increase. Figure \ref{fig:chns_valuediff} shows the difference in magnitude between selected pairwise differences in value function results between the Q-learning treatment rules,  assigning all individuals to either treatment $A = 1$, and treatment $A = 0$. \footnote{The complete pairwise comparison of all the comparisons is depicted in the appendix Figure \ref{fig:chns_valuediffsupp}}. Across all instances, the Q-learning models produced the most effective ITRs, with values decreasing significantly as $\delta_t$ increased. This trend indicates a reduced risk of hypertension in the heterogeneous, treatment-effect-augmented population, especially compared to population-level interventions where all individuals are assigned to high physical activity. When $\delta=1$, we observe that the pattern of value functions bears resemblance to that of the original data across various methods. As the prescriptive treatment effect becomes increasingly evident, both Q-learning and D-learning demonstrate progressively superior performance compared to the 'one-size-fits-all' approaches. This underscores the importance of personalizing over population-level interventions when certain factors significantly affect individual responses to treatment. Applying these methods to existing rich, detailed observational data can provide an important insight into factors that might distinguish these subgroups of responders. Future work will build on these existing models to better incorporate additional high-dimensional data. However, these results suggest that if a large heterogeneous effect truly exists, our method can successfully capture it.

\section{Discussion} \label{sec_discussion}
In this paper, we propose an innovative t-test based approach that can directly compare the value functions of any two treatment regimes. The validity of the approach follows from the asymptotic normality of the standard value function for the estimated treatment regime and the asymptotic normality of the propensity score model parameters. Our method provides valid estimates for (1) the variance and 95\% confidence interval for a value function for a single treatment regime, (2) the variance and 95\% confidence interval for the difference between two value functions of two treatment regimes, (3) the p-value of the t-test for the significance of the difference between two value functions, and (4) the application of these estimations in scenarios involving multiple imputations. Our method's estimates exhibit comparable Type I error and power performance to those obtained via bootstrap. Moreover, our approach simplifies variance calculation and is computationally more efficient, especially when multiple imputation is used to handle missing data. Through simulation studies and the data application example, we demonstrate the performance and the ease of implementation of our method in different scenarios. Additionally, this method enables the evaluation and comparison of ITR effectiveness using abundant observational cohort data when clinical trial data are not available. This comparison is crucial as individuals seek behavioral modification guidance amidst numerous recommendations, and emphasizing targeting strategies could increase the likelihood of effective changes.

Nevertheless, it is important to acknowledge a limitation inherent in our methodology – its reliance on the presupposition that the propensity score model holds true. In observational data, extreme propensity scores can influence the estimates of the value function as well as the variance estimates of the value function. For future work, it will be interesting to use more robust estimators to alleviate the impact of potential misspecification of the propensity score model. Another extension of this method is to multiple-stage ITR comparison. Extending our method to simultaneously compare more than two ITRs presents an interesting avenue for future research, which would require establishing the asymptotic normality of the joint propensity scores for multiple treatment assignments. 

To conclude, our method offers a convenient approach to comparing two treatment regimes directly. Furthermore, it is suitable for observational data and randomized trial data and has the ability to incorporate multiple imputation for missing data.


\section*{Acknowledgments}
This work was supported by the NIH, Eunice Kennedy Shriver National Institute of Child Health and Human Development (R01 HD30880), and the National Institute on Aging (R01AG065357). This research uses data from China Health and Nutrition Survey (CHNS). We are grateful to research grant funding from the National Institute for Health (NIH), the Eunice Kennedy Shriver National Institute of Child Health and Human Development (NICHD) for R01 HD30880 and R01 HD38700, National Institute on Aging (NIA) for R01 AG065357, National Institute of Diabetes and Digestive and Kidney Diseases (NIDDK) for R01 DK104371 and P30 DK056350, National Heart, Lung, and Blood Institute (NHLBI) for R01 HL108427, the NIH Fogarty grant D43 TW009077, the Carolina Population Center for P2C HD050924 and P30 AG066615 since 1989, and the China-Japan Friendship Hospital, Ministry of Health for support for CHNS 2009, Chinese National Human Genome Center at Shanghai since 2009, and Beijing Municipal Center for Disease Prevention and Control since 2011. We thank the National Institute for Nutrition and Health, China Center for Disease Control and Prevention, Beijing Municipal Center for Disease Control and Prevention, and the Chinese National Human Genome Center at Shanghai. In addition, Minxin Lu was supported by NC TraCS collaboration on "Optimizing weight status based on potentially modifiable risk factors". Both Minxin Lu and Michael Kosorok were funded in part by grant UM1 TR004406 from the National Center for Advancing Translational Sciences. We thank Matthew Christopher Brown and Lina Maria Montoya for their help and support for the project. The content is solely the responsibility of the authors and does not necessarily represent the official views of the NIH.

\subsection*{Author contributions}

Minxin Lu, Annie Green Howard, Penny Gordon-Larsen, Katie A. Meyer, Shufa Du, and Michael R. Kosorok contributed to the study conception and design. Data preparation was performed by Annie Green Howard and Hsiao-Chuan Tien. Data collection was performed by Huijin Wang and Bing Zhang. Statistical analysis was performed by Minxin Lu and Michael R. Kosorok. The first draft of the manuscript was written by Minxin Lu and all authors commented on previous versions of the manuscript. All authors read and approved the final manuscript.

\subsection*{Financial disclosure}

None reported.

\subsection*{Conflict of interest}

The authors declare no potential conflict of interests.

\clearpage

\bibliographystyle{unsrtnat}
\bibliography{references}  






\clearpage

\appendix
\renewcommand{\thetable}{S\arabic{table}}
\setcounter{table}{0}

\renewcommand{\thefigure}{S\arabic{figure}}
\setcounter{figure}{0}

\section{Proofs\label{app1}}
\textbf{Proof of Equation (\ref{value_derivation})}
\begin{proof}
$\sqrt{m}(\hat{V}_{m}(\hat{d}_{j,n})-V_0(\hat{d}_{j,n}))=$

$$\sqrt{m}(\frac{\sum_{i=1}^{m} \frac{y_i 1\{a_i=\hat{d}_{j,n}(x_i)\}}{\hat{\pi}_n(a_i|x_i)}-m E(\frac{Y1\{A=\hat{d}_{j,n}(X)\}}{\hat{\pi}_n(A|X)})}{\sum_{i=1}^{m}\frac{1\{a_i=\hat{d}_{j,n}(x_i)\}}{\hat{\pi}_n(a_i|x_i)}})$$

$$-\sqrt{m}(\frac{
    E[\frac{Y 1\{A=\hat{d}_{j,n}(X)\}}{\hat{\pi}_n(A|X))}]
    [\sum_{i=1}^{n}\frac{1\{a_i=\hat{d}_{j,n}(x_i)\}}{\hat{\pi}_n(a_i|x_i)}-m E(\frac{1\{A=\hat{d}_{j,n}(X)\}}{\hat{\pi}_n(A|X)})]}
    {E[\frac{1\{A=\hat{d}_{j,n}(X)\}}{\hat{\pi}_n(A|X))}]
    [\sum_{i=1}^{n}\frac{1\{a_i=\hat{d}_{j,n}(x_i)\}}{\hat{\pi}_n(a_i|x_i)}]})$$

$$+\sqrt{m}\left(\frac{E[\frac{Y 1\{A=\hat{d}_{j,n}(X)\}}{\hat{\pi}_n(A|X))}]-E[\frac{Y 1\{A=\hat{d}_{j,n}(X)\}}{\pi_0(A|X))}]}{E[\frac{1\{A=\hat{d}_{j,n}(X)\}}{\hat{\pi}_n(A|X))}]}\right)$$

$$-\sqrt{m}(\frac{
    E[\frac{Y 1\{A=\hat{d}_{j,n}(X)\}}{\pi_0(A|X))}]
    [E[\frac{1\{A=\hat{d}_{j,n}(X\}}{\hat{\pi}_n(A|X)}-E(\frac{1\{A=\hat{d}_{j,n}(X)\}}{\pi_0(A|X)})]}
    {E[\frac{1\{A=\hat{d}_{j,n}(X)\}}{\hat{\pi}_n(A|X))}]
    E[\frac{1\{A_=\hat{d}_{j,n}(X)\}}{\pi_0(A|X)}]})$$

    $$ = B_{1,m}-B_{2,m}+B_{3,m}-B_{4,m},$$

where 
    $$B_{1,m} = \sqrt{m}(\frac{\sum_{i=1}^{m} \frac{y_i 1\{a_i=\hat{d}_{j,n}(x_i)\}}{\hat{\pi}_n(a_i|x_i)}-m E(\frac{Y1\{A=\hat{d}_{j,n}(X)\}}{\hat{\pi}_n(A|X)})}{\sum_{i=1}^{m}\frac{1\{a_i=\hat{d}_{j,n}(x_i)\}}{\hat{\pi}_n(a_i|x_i)}})$$
    
    $$=\frac{\mathbb{G}_{n}(\frac{Y1\{A=\hat{d}_{j,n}(X)\}}{\hat{\pi}_n(A|X)})}
    {\mathbb{P}_{n}[\frac{1\{A=\hat{d}_{j,n}(X)\}}{\hat{\pi}_n(A|X)}]} = \mathbb{G}_{n} [\frac{Y1\{A=\hat{d}_{j,n}(X)\}}{\pi_0(A|X)}]+o_P(1),$$
where $\mathbb{G}_{n}((f(U)) = \sqrt{n}[\mathbb{P}_{n}(f(U))-P(f(U))]$, $\mathbb{P}_{n}f(U)=n^{-1}\sum_{i=1}^{n}f(u_i)$ is the empirical measure, $Pf(U)=E(f(U))$ is the expectation taken over $U$. By empirical process methods, $\mathbb{P}_n[\frac{1\{A=\hat{d}_{j,n}(X)\}}{\hat{\pi}_n(A|X)}] \rightarrow E[\frac{1\{A=\hat{d}_{j,n}(X)\}}{\pi_{0}(A|X)}]=1$, in probability. Using Slutsky's theorem, and the empirical average converges to its limiting value.\\

Next,
 $$B_{2,m}=\frac{E[\frac{Y1\{A=\hat{d}_{j,n}(X)\}}{\hat{\pi}_n(A|X)}] \mathbb{G}_{n}[\frac{1\{A=\hat{d}_{j,n}(X)\}}{\hat{\pi}_n(A|X)}]}
    {E[\frac{1\{A=\hat{d}_{j,n}(X)\}}{\hat{\pi}_n(A|X)}] \mathbb{P}_{n}[\frac{1\{A=\hat{d}_{j,n}(X)\}}{\hat{\pi}_n(A|X)}]} 
    =  V_0(\hat{d}_{j,n}) \mathbb{G}_{n} (\frac{1\{A=\hat{d}_{j,n}(X)\}}{\pi_0(A|X)})+o_P(1),$$
    by the following steps based on empirical process methods and Slutsky's theorem:\\

\begin{align*} 
& \frac{E[\frac{Y1\{A=\hat{d}_{j,n}(X)\}}{\hat{\pi}_n(A|X)}]}{E[\frac{1\{A=\hat{d}_{j,n}(X)\}}{\hat{\pi}_n(A|X)}]} 
     =\frac{E[\frac{Y1\{A=\hat{d}_{j,n}(X)\}}{\pi_0(A|X)}]}{E[\frac{1\{A=\hat{d}_{j,n}(X)\}}{\pi_0(A|X)}]}+o_P(1)
    =V_0(\hat{d}_{j,n}) +o_P(1), \\
 & \mathbb{P}_{n}[\frac{1\{A=\hat{d}_{j,n}(X)\}}{\hat{\pi}_n(A|X)}] =1+o_P(1),\\
 & \mathbb{G}_{n}[\frac{1\{A=\hat{d}_{j,n}(X)\}}{\hat{\pi}_n(A|X)}] = \mathbb{G}_{n}[\frac{1\{A=\hat{d}_{j,n}(X)\}}{\pi_0(A|X)}] +o_P(1).
\end{align*}
\\
Next,
     
    $$B_{3,m}=\sqrt{m}\left(\frac{E[\frac{Y 1\{A=\hat{d}_{j,n}(X)\}}{\hat{\pi}_n(A|X))}]-E[\frac{Y 1\{A=\hat{d}_{j,n}(X)\}}{\pi_0(A|X))}]}{E[\frac{1\{A=\hat{d}_{j,n}(X)\}}{\hat{\pi}_n(A|X))}]}\right)$$
    $$=\frac{E[Y1\{A=\hat{d}_{j,n}(X)\}] E[\sqrt{m}(\frac{1}{\hat{\pi}_n(A|X)}-\frac{1}{\pi_0(A|X)})]}
    {E[\frac{1\{A=\hat{d}_{j,n}(X)\}}{\hat{\pi}_n(A|X)}]}$$

    $$= - E[\frac{Y1\{A=\hat{d}_{j,n}(X)\}\sqrt{m}(\hat{\pi}_n(A|X)-\pi_0(A|X))}{\pi^2_0(A|X)}]+o_P(1).$$
  
    Since $\sqrt{n}(\pi(a|x,\hat{\theta}_n)-\pi(a|x,\theta_0))=\sqrt{n}(\hat{\theta}_n-\theta_0)^T\phi_0(a,x)+o_p(1)$, we now have that
    
    $$B_{3,m}=-\sqrt{m/n}*\sqrt{n}(\hat{\theta}_n-\theta_0)^T E [\phi_0(A,X)\frac{Y1\{A=\hat{d}_{j,n}(X)\}}{\pi_0^2(A|X)}]+o_P(1).$$

Finally,
    $$B_{4,m}=-\sqrt{m}\frac{
    E(\frac{Y1\{A=\hat{d}_{j,n}(X)\}}{\pi_0(A|X)})
    E[1\{A=\hat{d}_{j,n}(X)\}(\frac{1}{\hat{\pi}_n(A|X)}-\frac{1}{\pi_0(A|X)})]}
    {E(\frac{1\{A=\hat{d}_{j,n}(X)\}}{\pi_0(A|X)})
    E(\frac{1\{A=\hat{d}_{j,n}(X)\}}{\hat{\pi}_n(A|X)})}$$

    $$=-V_0(\hat{d}_{j,n})E[1\{A=\hat{d}_{j,n}(X)\}\sqrt{m}(\frac{1}{\hat{\pi}_n(A|X)}-\frac{1}{\pi_0(A|X)})]+o_P(1)$$

    $$=V_0(\hat{d}_{j,n})E[\frac{1\{A=\hat{d}_{j,n}(X)\}}{\pi_0^2(A|X)}\sqrt{m}(\hat{\pi}_n(A|X)-\pi_0(A|X))]+o_P(1)$$
    $$=\sqrt{m/n}\sqrt{n}(\hat{\theta}_n-\theta_0)^T
    E[\phi_0(A,X)\frac{1\{A=\hat{d}_{j,n}(X)\}}{\pi_0^2(A|X)}]V_0(\hat{d}_{j,n})+o_P(1).$$
    
Thus:
\begin{align*}
    \sqrt{m}(\hat{V}_{m}(\hat{d}_{j,n})-V_0(\hat{d}_{j,n}))& = 
    \mathbb{G}_{m}(\frac{Y1\{A=\hat{d}_{j,n}(X)\}}{\pi_0(A|X)})\\
    & -V_0(\hat{d}_{j,n})\mathbb{G}_{m}(\frac{1\{A=\hat{d}_{j,n}(X)\}}{\pi_0(A|X)})\\
    & -\sqrt{m/n}\sqrt{n}(\hat{\theta}_n-\theta_0)^T
    E[\phi_0(A|X)\frac{Y1\{A=\hat{d}_{j,n}(X)\}}{\pi^2_0(A|X)}]\\
    & +\sqrt{m/n}\sqrt{n}(\hat{\theta}_n-\theta_0)^T
    E[\phi_0(A|X)\frac{1\{A=\hat{d}_{j,n}(X)\}}{\pi^2_0(A|X)}]V_0(\hat{d}_{j,n})+o_P(1)\\
    & = \mathbb{G}_{m}(\frac{(Y- V_0(\hat{d}_{j,n}))1\{A=\hat{d}_{j,n}(X)\}}{\pi_0(A|X)})\\
    &-\sqrt{m/n}\sqrt{n}(\hat{\theta}_n-\theta_0)^T
    E(\phi_0(A,X)\frac{(Y-V_0(\hat{d}_{j,n}))1\{A=\hat{d}_{j,n}(X)\}}{\pi^2_0(A|X)})+o_P(1),
\end{align*}
and the desired results follow.
\end{proof}

\textbf{Proof of Proposition 2}
\begin{proof}
$$\sqrt{m}(\hat{V}_{m}(\hat{d}_{1,n})-\hat{V}_{m}(\hat{d}_{2,n})-V_0(\hat{d}_{1,n})+V_0(\hat{d}_{2,n}))$$

$$= \mathbb{G}_{m}[\frac{(Y-V_0(\hat{d}_{1,n}))1\{A=\hat{d}_{1,n}(X)\}-(Y-V_0(\hat{d}_{2,n}))1\{A=\hat{d}_{2,n}(X)\}
}{\pi_0(A|X)}]$$
    
$$ -\sqrt{m/n}\sqrt{n}(\hat{\theta}_n-\theta_0)^T
E(\phi_0(A,X)\frac{(Y-V_0(\hat{d}_{1,n}))1\{A=\hat{d}_{1,n}(X)\}-(Y-V_0(\hat{d}_{2,n}))1\{A=\hat{d}_{2,n}(X)\}}
{\pi_0^2(A|X)})
+o_P(1)
$$

$$\rightarrow N(0,T_{0}^2),$$
in distribution, where

\begin{equation*}
    T_{0}^2=Var(U_1-U_2)
    +\frac{m}{n}(W_{1}-W_{2})^T \Sigma_{0}(W_{1}-W_{2}),
\end{equation*}

and standard arguments can now be used to show that  $\hat{T}_{m}^2 \rightarrow T_{0}^2$ as $m\rightarrow \infty$, where
$$\hat{T}_{m}^2=m^{-1}\sum_{i=1}^{m}(\hat{U}_{i,1}-\hat{U}_{i,2}-\bar{U}_{1}+\bar{U}_{2})^2+\frac{m}{n}(\hat{W}_{1}-\hat{W}_{2})^T\hat{\Sigma}_{n}(\hat{W}_{1}-\hat{W}_{2}).$$

\end{proof}

\textbf{Proof of Assumptions: logistic regression as an example}
\begin{proof}
In general, most generalized linear models satisfy all assumptions under standard regularity conditions. Here, we illustrate this using logistic regression as an example. The first assumption is automatically satisfied for logistic regression. Suppose the propensity score model is specified as $\pi(a|x, \theta_0) = \frac{e^{a\theta_0^T x}}{1 + e^{\theta_0^T x}}$, for $a \in \{0, 1\}$. 

The left-hand side of Equation (1) becomes 
\[
\sqrt{n} \left( \frac{e^{a \hat{\theta}_n^T x}}{1 + e^{\hat{\theta}_n^T x}} - \frac{e^{a \theta_0^T x}}{1 + e^{\theta_0^T x}} \right).
\]

Using a Taylor expansion of $\hat{\theta}_n$ around $\theta_0$ for both $a = 1$ and $a = 0$, we obtain:
\[
\sqrt{n}(\hat{\theta}_n - \theta_0)^T (2a - 1) x \cdot \frac{e^{a \theta_0^T x}}{(1 + e^{\theta_0^T x})^2} + o_P(1).
\]

Therefore, we define
\[
\phi_0(a, x) = x(2a - 1) \cdot \frac{e^{a \theta_0^T x}}{(1 + e^{\theta_0^T x})^2}, \quad \text{and} \quad \hat{\phi}_n(a, x) = x(2a - 1) \cdot \frac{e^{a \hat{\theta}_n^T x}}{(1 + e^{\hat{\theta}_n^T x})^2}.
\]
This confirms that Assumption 1 is satisfied.

Assumption 2 holds by the asymptotic normality of the maximum likelihood estimator (MLE) for logistic regression, provided the standard covariance estimator from the MLE is used. An implication of this assumption is that the propensity score model must be correctly specified.

We can also verify that the third assumption holds. From Assumption 2, we have that $\hat{\theta}_n \rightarrow \theta_0$ in probability as $n \rightarrow \infty$. By the continuous mapping theorem, it follows that $\hat{\phi}_n(A, X) \rightarrow \phi_0(A, X)$ in probability as $n \rightarrow \infty$. 

Moreover, since $\|\hat{\phi}_n(A, X) - \phi_0(A, X)\| \leq \|X\|$ and $E\|X\|^2 < \infty$, we can apply the dominated convergence theorem to conclude that 
\[
P\|\hat{\phi}_n(A, X) - \phi_0(A, X)\|^2 \rightarrow 0,
\]
in probability as $n \rightarrow \infty$.

\end{proof}
\clearpage

\section{Additional Tables and Figures\label{app2}}%

\begin{center}
\begin{table*}[ht]%
\caption{Estimated value functions of different treatment regimes for CHNS data.\label{tab:chns_original_value}}
\centering
\resizebox{\textwidth}{!}{%
\begin{tabular*}{500pt}{@{\extracolsep\fill}lccD{.}{.}{3}c@{\extracolsep\fill}}
\toprule
&\multicolumn{2}{@{}c@{}}{\textbf{Training Set}} & \multicolumn{2}{@{}c@{}}{\textbf{Test Set}} \\\cmidrule{2-3}\cmidrule{4-5}
\textbf{} & \textbf{Mean}  & \textbf{SD}  & \multicolumn{1}{@{}l@{}}{\textbf{Mean}}  & \textbf{SD}   \\
\midrule
Obs & 0.602  & 0.021  &  0.595 & 0.031  \\
AllLowPA & 0.603  & 0.018  & 0.591  & 0.026   \\
AllHighPA & 0.600  & 0.011  &  0.598 & 0.017   \\
Q-Learning & 0.574  & 0.013  & 0.573  & 0.017  \\
D-Learning & 0.576  & 0.016  &  0.598 & 0.023  \\
\bottomrule
\end{tabular*}
}
\end{table*}
\end{center}

\begin{center}
\begin{table*}[ht]
\caption{Estimated value differences and p-values of selected comparisons between treatment regimes for CHNS data.\label{tab:chns_original_valuediff}}
\centering
\resizebox{\textwidth}{!}{%
\begin{tabular*}{500pt}{@{\extracolsep\fill}lcccccc}
\toprule
& \multicolumn{3}{c}{\textbf{Training Set}} & \multicolumn{3}{c}{\textbf{Test Set}} \\
\cmidrule{2-4}\cmidrule{5-7}
\textbf{Difference} & \textbf{Mean} & \textbf{SD} & \textbf{P-Value} & \textbf{Mean} & \textbf{SD} & \textbf{P-Value} \\
\midrule
AllLowPA vs. Obs & 0.001 & 0.011 & 0.924 & -0.004 & 0.017 & 0.827 \\
AllHighPA vs. Obs & -0.001 & 0.018 & 0.953 & 0.003 & 0.026 & 0.897 \\
AllHighPA vs. AllLowPA & -0.002 & 0.021 & 0.921 & 0.007 & 0.032 & 0.821 \\
Q vs. Obs & -0.028 & 0.017 & 0.096 & -0.022 & 0.026 & 0.407\\
Q vs. AllLowPA & -0.029 & 0.017 & 0.086 & -0.018 & 0.026 & 0.497 \\
Q vs. AllHighPA & -0.027 & 0.013 & 0.043 & -0.025 & 0.018 & 0.156 \\
Q vs. D & -0.002 & 0.013 & 0.871 & -0.024 & 0.020 & 0.223 \\
\bottomrule
\end{tabular*}
}
\end{table*}
\end{center}


\begin{figure}[ht]\centerline{\includegraphics[width=450pt,height=300pt]{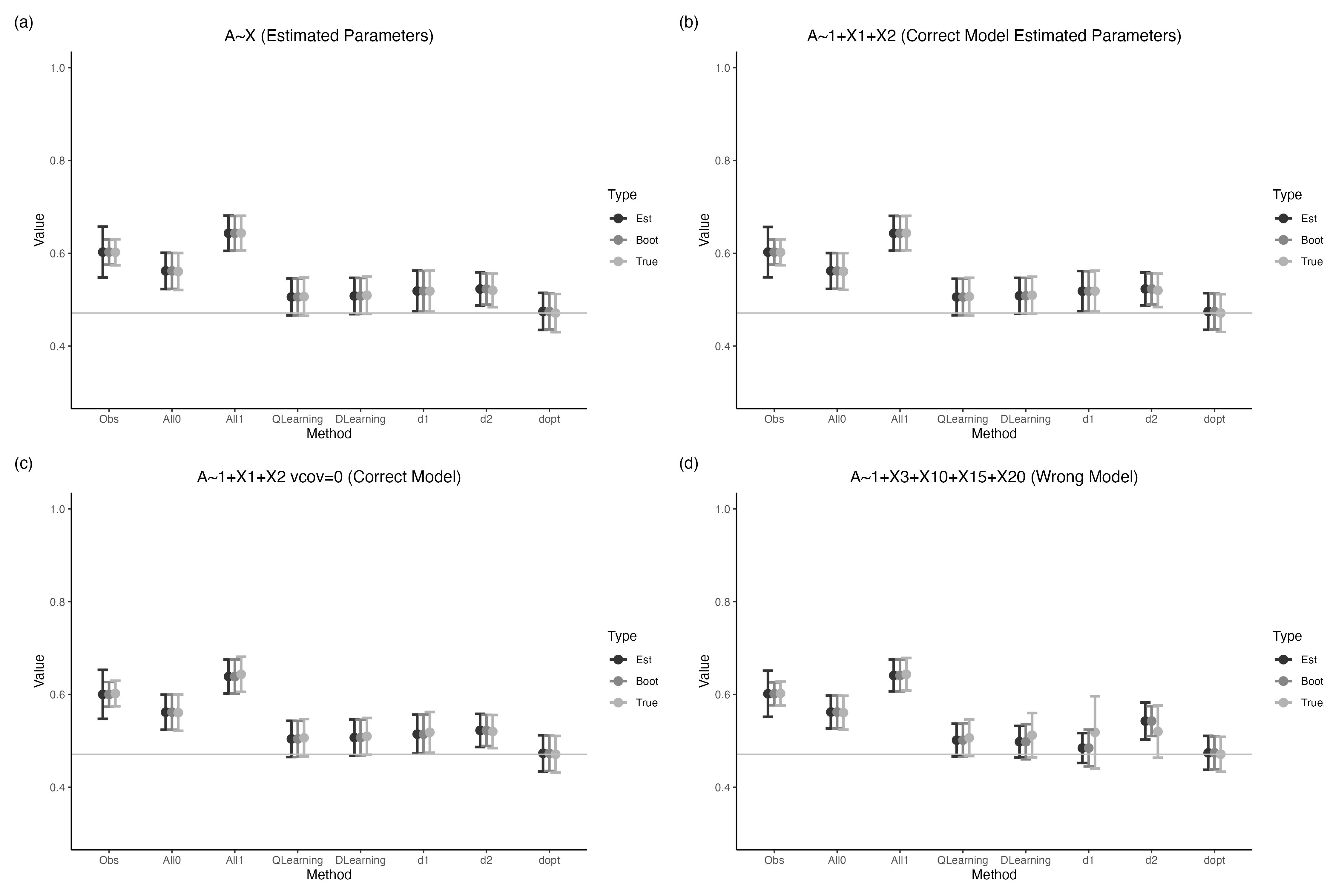}}
\caption{Value functions and 95\% confidence interval estimates obtained using propensity scores obtained from different propensity score models: (a) a logistic regression model including all covariates $\hat{\pi}= \hat{\theta} X$, (b) a correctly specified model containing only truly relevant covariates with estimated parameters $\hat{\pi}= \hat{\theta}^\top [1, X_1, X_2]$, (c) the true propensity score model with known parameters $\pi_0= \theta_0^\top [1, X_1, X_2]$, and (d) an incorrectly specified model including only non-relevant covariates with estimated parameters $\hat{\pi}= \hat{\theta}^\top [1, X_3, X_{10},X_{15},X_{20}]$. Results shown here are for scenario 4 (observational data with data missing at random) with $n=5000$ training samples, and $\delta_t=1$. Scenario 3 shows similar results and is thus omitted. When the propensity score is incorrect, the true value function remains the same for $d_1$ $d_2$ $d_{opt}$, the estimates have larger bias, and thus the true variance becomes larger. The true variance for value difference equals the average of the squared difference between the estimated value difference and the true value difference.}
\label{fig:sim_value_propen}
\end{figure}

\begin{figure}[ht]\centerline{\includegraphics[width=450pt,height=300pt]{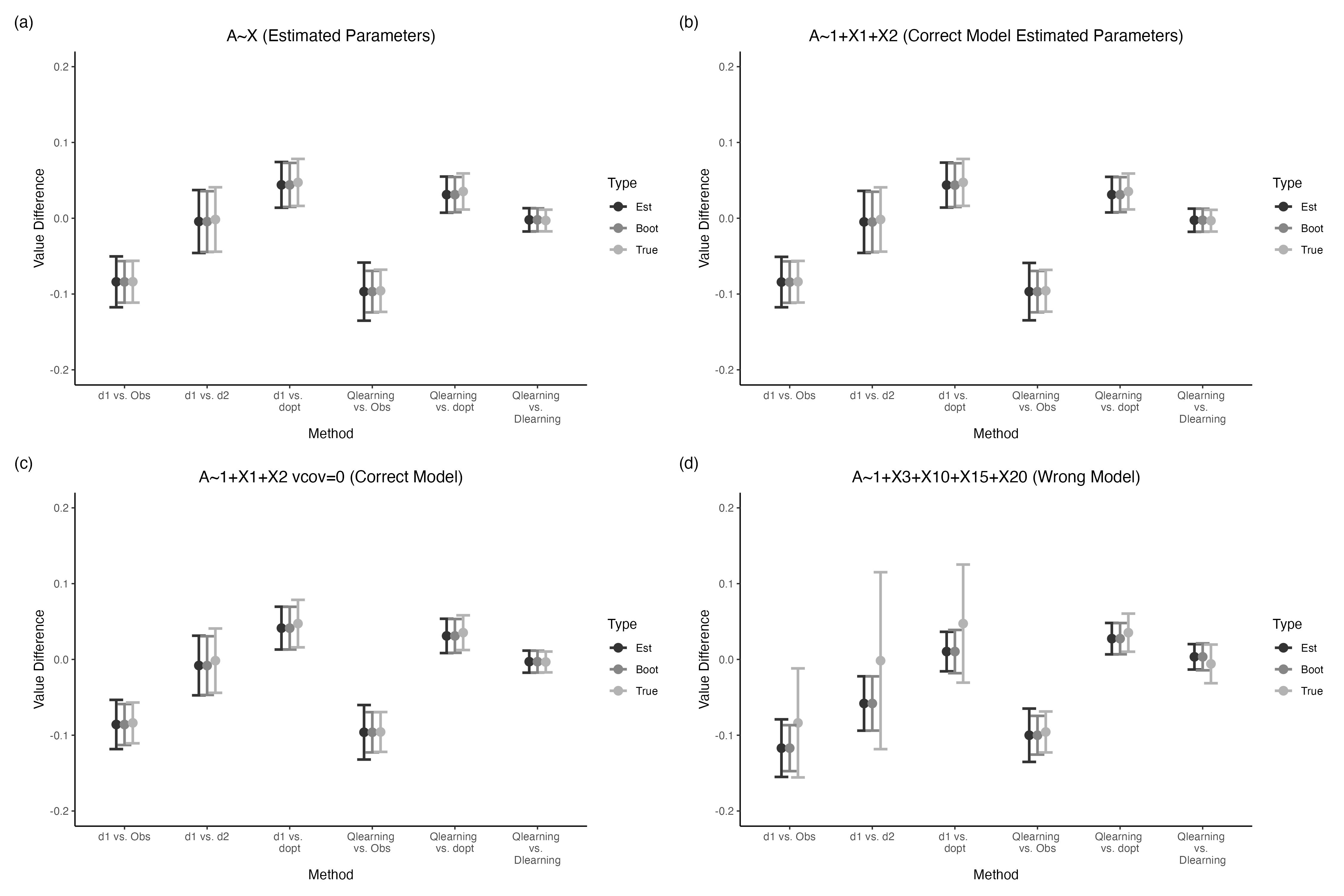}}
\caption{Differences between the two value functions and their corresponding 95\% confidence interval estimates, obtained using propensity scores derived from different models: (a) a logistic regression model including all covariates $\hat{\pi}= \hat{\theta} X$, (b) a correctly specified model containing only truly relevant covariates with estimated parameters $\hat{\pi}= \hat{\theta}^\top [1, X_1, X_2]$, (c) the true propensity score model with known parameters $\pi_0= \theta_0^\top [1, X_1, X_2]$, and (d) an incorrectly specified model including only non-relevant covariates with estimated parameters $\hat{\pi}= \hat{\theta}^\top [1, X_3, X_{10},X_{15},X_{20}]$. Results shown here are for scenario 4 with n=5000 training samples and $\delta_t=1$. Scenario 3 shows similar results and is thus omitted. }
\label{fig:sim_valuediff_propen}
\end{figure}
\clearpage
\begin{figure}[ht]
\centering\centerline{\includegraphics[width=342pt,height=200pt]{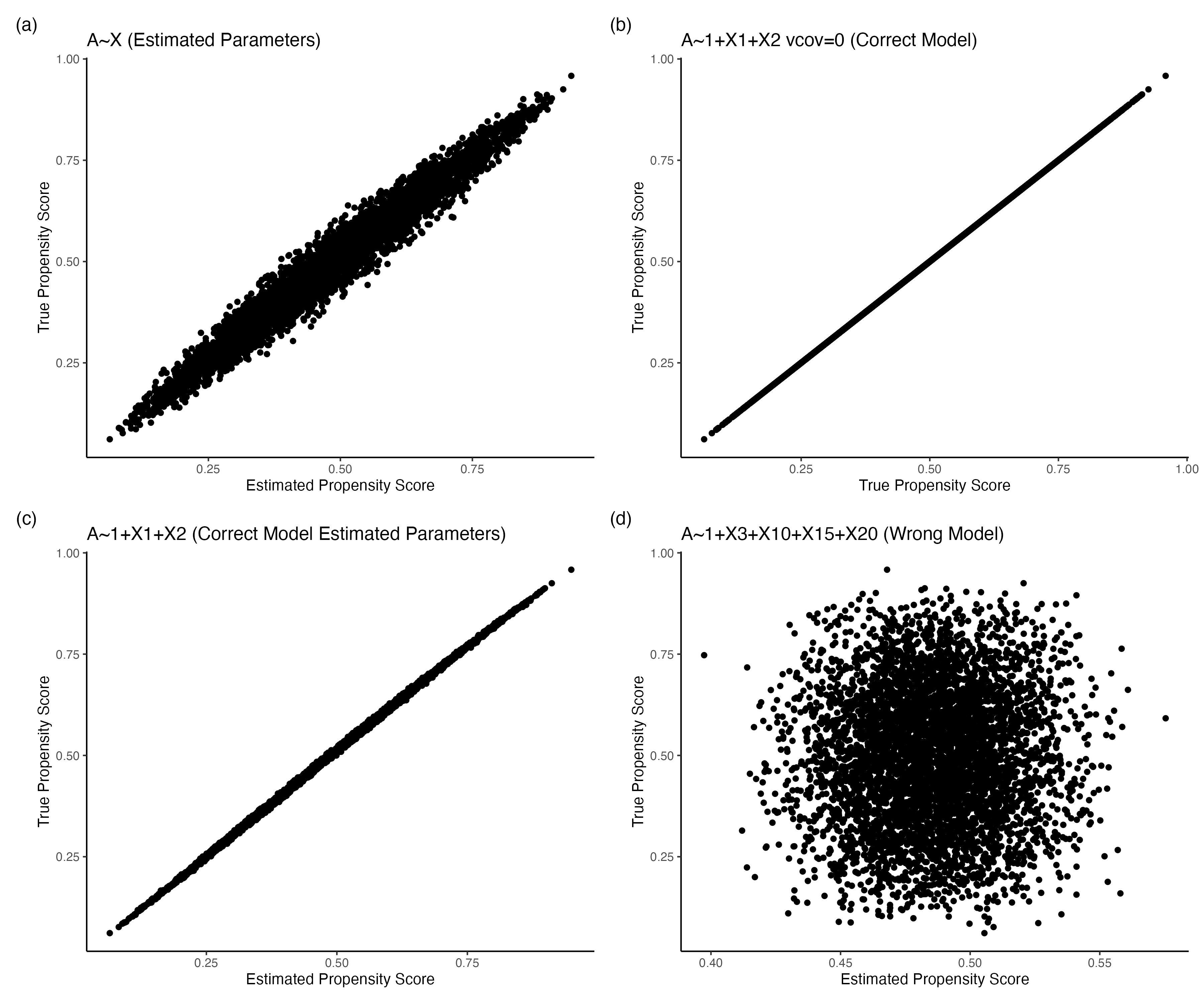}}
    \caption{A demonstration of one estimated propensity score compared to the true propensity score in propensity score model misspecification sensitivity analysis in one replication.}
    \label{fig:propen_demo}
\end{figure}

\clearpage

\begin{figure}[ht]
\centering\centerline{\includegraphics[width=342pt,height=200pt]{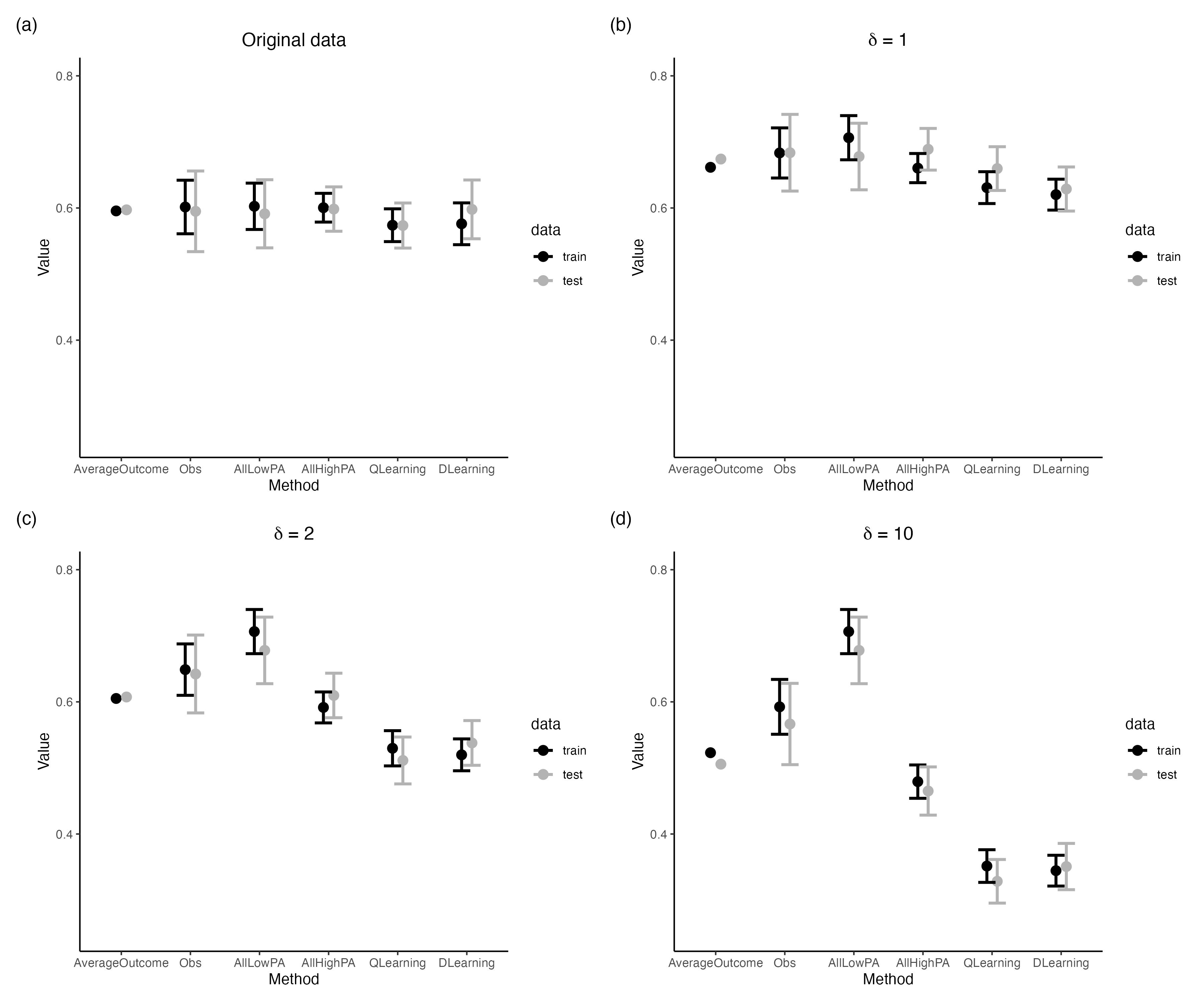}}
\caption{The value function results for both training and test sets for five different treatment regimes: observed treatment, treatment $A=0$ for all individuals, treatment $A=1$ for all individuals, Q-learning optimal ITR, and D-learning optimal ITR. For comparison purposes, we include the average outcome as a benchmark. The results on original data (a) and three simulated data sets with three levels of treatment effect modification are shown (b) $\delta=1$, (c) $\delta=2$, (d) $\delta=10$. }
\label{fig:chns_valuesupp}
\end{figure}

\clearpage

\begin{figure}[ht]
\centering
\includegraphics[width=342pt,height=200pt]{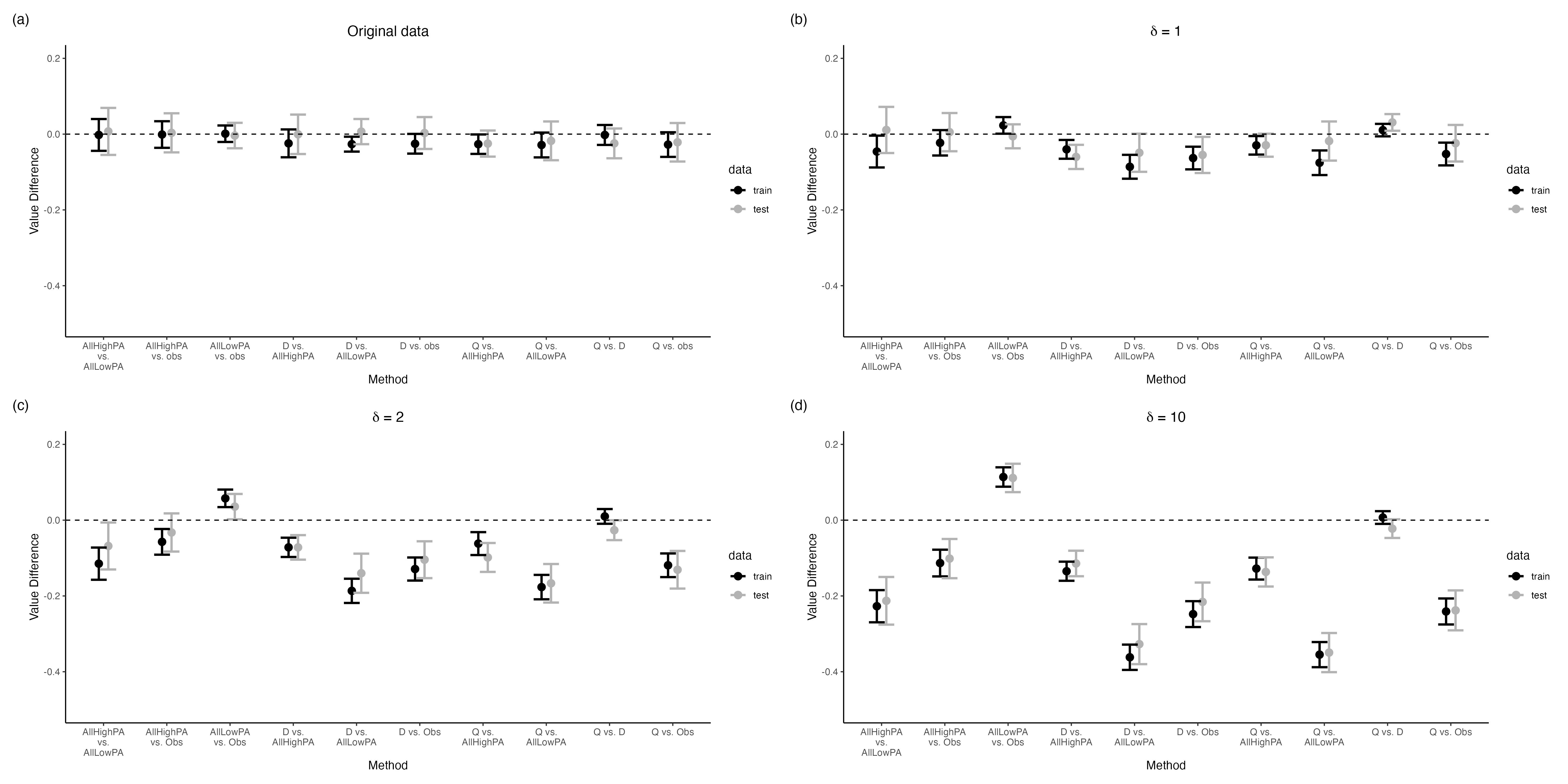}
\caption{The differences of value functions between the pairs of the following treatments for both training and test sets:  treatment $A=0$ for all individuals, treatment $A=1$ for all individuals, Q-learning optimal ITR, and D-learning optimal ITR. The results on original data (a) and three simulated data with three levels of treatment effect modification are shown (b) $\delta=1$, (c) $\delta=2$, (d) $\delta=10$.
\label{fig:chns_valuediffsupp}}
\end{figure}

\end{document}